\documentclass[10pt,a4paper]{article}

\usepackage{amsmath,amssymb,amsfonts}
\usepackage[dvips]{graphicx}
\usepackage{epsfig}
\usepackage{subcaption}
\usepackage[T1]{fontenc}
\usepackage{amsmath}
\usepackage{amsfonts}
\usepackage{amssymb}
\usepackage{color}
\usepackage{graphicx}
\usepackage{caption}
\usepackage{subcaption}
\usepackage{amsmath,amssymb,amsfonts,graphicx}
\usepackage{epsfig}

\newcommand{\beq}{\begin{eqnarray}}
	\newcommand{\eeq}{\end{eqnarray}}

\usepackage[left=2.4cm,top=3.3cm,right=2.4cm,bottom=3.3cm,bindingoffset=0cm]{geometry}

\begin{document}

	\title{ 
		\vskip 30pt
		Code generation for AMReX with applications to numerical relativity}
	
	\vskip 30pt 
	
	\author{Adam J Peterson$^a$, Don Willcox$^a$, \\ and Philipp M\"osta$^b$
		\vspace{0 cm}\\
		\normalsize {\it $^{a}$ Center for Computational Science and Engineering}, \\
        {\it Lawrence Berkeley National Laboratory}, Berkeley, CA 94720 USA
		\vspace{.3 cm}\\
		\normalsize \it $^{b}$ GRAPPA, Anton Pannekoek Institute for Astronomy,\\
            \normalsize \it Institute of High-Energy Physics, and Insitute of Theoretical Physics\\
            \normalsize \textit{University of Amsterdam}, Science Park 904, 1098 XH Amsterdam, The Netherlands
		\vspace{.3 cm}\\
		{ \footnotesize ajpeterson@lbl.gov}  
	}

	\date{\hfill}
	\maketitle
	
	\begin{abstract}
		
		We present a new python/SymPy based code generator for producing executable numerical expressions for partial differential equations in AMReX-based applications.  We demonstrate the code generator capabilities for the case of $3+1$ ADM formulations of numerical relativity for the constraint damped, conformal Z4 formulations (Z4c and CCZ4).  The generated spacetime solvers are examined for stability and accuracy using a selection of checks from the standard Apples with Apples testbeds for numerical relativity applications.  We also explore physically interesting vacuum spacetimes including head-on and inspiraling black hole binary collisions, and investigate the simulated gravitational waveforms from such events with the Newman-Penrose formulation of waveform extraction.

	\end{abstract}

\newpage


\section{Introduction}

One of the most exciting developments in astrophysics and cosmology in the last decade has been the arrival of gravitational wave astronomy.  Observations from LIGO \cite{Abramovici:1992ah}, VIRGO \cite{Caron:1997hu}, and other gravitational wave observatories have opened up a new probe into the universe \cite{LIGOScientific:2016aoc}, and in particular, new testing into the strong field limit of general relativity via black hole/neutron star collisions.  For both observational and testing purposes there comes the necessity to accurately simulate relativistic spacetime physics in the regions of large curvature (non-perturbative), such as those in the vicinity of black holes and neutron stars.  Also of interest is the dynamical evolution of stellar core collapse supernovae, which for proper understanding, requires accurate evolution of general relativistic/magnetohydrodynamic systems.  

For these reasons (and many others) a market has developed for large scale numerical simulations capable of evolving general relativistic systems in both vacuum and matter filled spacetimes.  In the past decade many codes have been developed that successfully evolve spacetimes for both black hole and neutron star systems, in particular those in compact binary evolutions.  It has also become clear that such complex systems involve highly non-trivial partial differential equations associated with the 3+1 decomposition of Einstein's equations.  Since no single code can capture all of the physics of interest for all studies of astrophysical and cosmological systems, in many cases one is required to develop codes from scratch that are tailored for the particular physics of interest.  For each study, one is required to translate tensor symbolic expressions to computational numerical language (finite-differencing, index expansions, etc.) for each separate project. For such complex numerical systems, this is both impractical and inconvenient for the researcher interested in modeling multiple physical phenomena in the context of high spacetime curvature.

It has also become clear in the past few years that $3+1$ formulations of General Relativity provide excellent testing grounds for new PDE code generators, as they are well studied systems with numerous examples in the literature to compare results. Formulations such as the BSSNOK \cite{Baker:2005vv} \cite{Campanelli:2005dd} and Z4 \cite{Bona:2003fj} \cite{Gundlach:2005eh} \cite{Bernuzzi:2009ex} approaches provide particularly rigorous tests of code generation due to their large number of evolved variables, non-linearity of their equations of motion, and multiple finite-difference stenciling for both forward, backward, and centered indexing.  They are also known to be numerically stable \cite{Gundlach:2006tw} if formulated correctly, and thus provide a sensitive test of the correctness of the generated code \cite{Babiuc:2007vr}.

The goal of this project is to develop and demonstrate the utility of code generation to facilitate the translation of symbolic tensor formulations of equations to discretized formulas written in low level programming languages such as C++ or Fortran.  Python based symbolic manipulation tools such as SymPy, NumPy, and SciPy are excellent applications for performing calculations on symbolic/tensor objects.  However, as numerical solvers for large systems increasingly rely on parallelization and AMR based applications that are typically written in low level languages, it is necessary to develop translators that can generate executable expressions for such languages.  

To this end we develop a code generator capable of producing executable expressions for AMReX based PDE solvers from symbolic manipulation tools written in SymPy and NumPy.  Previous projects have produced packages designed specifically for mathematica \cite{Husa:2004ip} and python \cite{Ruchlin:2017com} interfaces with more examples found in the Einstein Toolkit, and we wish to extend these interests to a python-to-AMReX code generation. Specifically, for this project, we demonstrate code generation for the Z4c and CCZ4 formulations of general relativity and evaluate the resulting system using selected examples of Apples with Apples tests for numerical spacetime solvers.  AMReX provides the necessary infrastructure for developing massively parallel block structured adaptive mesh refinement applications.  In particular, AMReX provides the necessary tools to evolve hyperbolic PDEs associated with 3+1 formulations of general relativity.  The code generator is designed to read textbook expressions of symbolic equations written in SymPy and produce the necessary finite differenced equivalents written as AMReX executable lines.

We organize our presentation as follows: We begin in section 2 with a review of the Z4c formulations of general relativity.  In section 3 discuss numerical methods for finite differencing and AMReX's algorithms for adaptive mesh refinement.  Section 4 is devoted to the specifics of code generation for AMReX applications. In section 5 we demonstrate the AwA tests for our numerical solvers, and apply the code to black hole binary systems in section 6.  We conclude with a discussion and summary in section 7.

\section{Review of $3+1$ formulation of Einstein equations}
For the bulk of the simulations performed in this project, we used the $3+1$ conformal Z4 (Z4c) formulation \cite{Bernuzzi:2009ex} of numerical relativity with constraint damping \cite{Bona:2003fj}.  The Z4c formulation is based on the extension of Einstein's equations to include constraint damping terms for stability purposes. More specifically, the Z4c formulation is given as follows:
\begin{align}
	G_{ab} &= 8\pi T_{ab} - 2\nabla_{(a} Z_{b)} + g_{ab} \nabla_c Z^c \nonumber \\
	&+ \kappa_1 \left[ 2 n_{(a} Z_{b)} + \kappa_2 g_{ab} n_c Z^c\right],
\label{Z4cEE}
\end{align}
with the Einstein tensor:
\begin{equation}
	G_{ab} = R_{ab} - \frac{1}{2} g_{ab} R.
\end{equation}
Here, $R_{ab}$ is the Ricci tensor and $R$ the Ricci scalar.  $\nabla_a$ is the covariant derivative compatible with the metric $g_{ab}$.  $Z_a$ are a vector field of constraints.  The fields $n_a$ are the timelike unit normals to the spacelike hypersurfaces in the $3+1$ decomposition of the spacetime manifold.  $Z_a$ are decomposed into $\Theta \equiv -n^a Z_a$, and $Z_i = \perp^a_i Z_a$. Clearly, if $Z_a = 0$, solutions to (\ref{Z4cEE}) are also solutions of the Einstein equations.

The introduction of the fields $Z_a$ in (\ref{Z4cEE}) introduce additional evolution equations in the Z4c system.  Thus in addition to the dynamical evolution of the spatial metric $\gamma_{ij}$ and extrinsic curvature $K_{ij}$,
\begin{align}
	\partial_t \gamma_{ij} &= -2\alpha K_{ij} + \mathcal{L}_\beta \gamma_{ij} \nonumber \\
	\partial_t K_{ij} &= - D_i D_j \alpha + \alpha \left[R_{ij} + KK_{ij} -2K_{ik}K^{k}_j + 2\hat{D}_{(i}Z_{j)} \right. \nonumber \\
	& \,\left. \phantom{\hat{D}} -\kappa_1(1+\kappa_2)\Theta\gamma_{ij}\right] + 4\pi \alpha\left[ \gamma_{ij}(S-\rho_{\rm ADM})-2S_{ij}\right] + \mathcal{L}_\beta K_{ij},
\label{Z4gammaK}
\end{align}
additional equations for $\Theta$ and $Z_i$ are included in the dynamical evolution:
\begin{align}
	\partial_t\Theta &= \frac{1}{2}\alpha \left[H + 2\hat{D}^i Z_i - 2 \kappa_1(2+\kappa_2)\Theta \right] + \mathcal{L}_\beta \Theta \nonumber \\
	\partial_t Z_i &= \alpha \left[M_i +D_i \Theta - \kappa_1 Z_i\right] + \gamma^{1/3} Z^j \partial_t\left[\gamma^{-1/3}\gamma_{ij} \right] + \beta^j \hat{D}_j Z_i,
\label{Z4ThetaZ}
\end{align}
with the definition
\begin{equation}
	\hat{D}_i Z_j \equiv \gamma^{1/3}\gamma_{kj}\partial_i \left[\gamma^{-1/3}Z^k\right].
\end{equation}
Here $\gamma \equiv \det{\gamma_{ij}}$.  Additionally, the Hamiltonian and momentum constraints are written as:
\begin{align}
	H &= R - K_{ij}K^{ij} + K^2 - 16 \pi \rho_{\rm ADM} \\
	M_i &= D^j\left[ K_{ij} - \gamma_{ij} K \right] - 8 \pi S_i .
\label{PhysicalConstraints}	
\end{align}

The Z4c system is further developed with a conformal transformation of the spatial metric, along with additional conformal transformations of the dynamical variables.  We define the following:

\begin{align}
	\tilde{\gamma}_{ij} &= \gamma^{-1/3}\gamma_{ij}, & 	\phi &= \frac{1}{12} \log\gamma \\
	\tilde{A}_{ij} &= \gamma^{-1/3}\left(K_{ij}-\frac{1}{3}\gamma_{ij}K\right), & \hat{K} &= K - 2 \Theta \\
	\tilde{\Gamma}^i &= 2 \tilde{\gamma}^{ij}Z_j + \tilde{\gamma}^{ij}\tilde{\gamma}^{kl}\partial_l \tilde{\gamma}_{jk}, & \tilde{\Gamma}_{\rm d}^i &= \tilde{\Gamma}^i_{jk}\tilde{\gamma}^{jk}
\end{align}

With these definitions the Z4 system (\ref{Z4gammaK}) and (\ref{Z4ThetaZ}) takes the form

\begin{align}
	\partial_t \phi &= -\frac{1}{6}\alpha (\hat{K} + 2 \Theta) + \beta^{i}\partial_{i}\phi + \frac{1}{6}\partial_i \beta^i \\
	\partial_t \tilde{\gamma}_{ij} &= -2 \alpha \tilde{A}_{ij} + 2 \tilde{\gamma}_{k(i}\partial_{j)}\beta^k -\frac{2}{3}\tilde{\gamma}_{ij}\partial_k \beta^k+ \beta^k \partial_k\tilde{\gamma}_{ij} \\
	\partial_t \hat{K} &= -D_i D^i \alpha + \alpha \left[ \tilde{A}_{ij} \tilde{A}^{ij} + \frac{1}{3}(\hat{K}+2\Theta)^2 \right. \nonumber \\
						&\;\;\;\;    \left. \vphantom{\frac{0}{0}} +\kappa_1(1-\kappa_2)\Theta \right] + 4 \pi \alpha \left(S +\rho_{\rm ADM}\right) + \beta^i \partial_i \hat{K} \\
	\partial_t \tilde{A}_{ij} &= e^{-4\phi}\left[ -D_i D_j \alpha + \alpha \left( R_{ij} - 8 \pi S_{ij}\right)\right]^{\rm tf} + \alpha \left[ \left( \hat{K} + 2\Theta \right) \tilde{A}_{ij} \right. \nonumber \\
	&\left. \;\;\; -2\tilde{A}_{ik}\tilde{A}^k_j\right] + 2\tilde{A}_{k (i}\partial_{j)}\beta^k-\frac{2}{3}\tilde{A}_{ij} \partial_k \beta^k+\beta^k \partial_k \tilde{A}_{ij} \\
	\partial_t \Theta &= \frac{1}{2}\alpha\left[R - \tilde{A}_{ij}\tilde{A}^{ij} + \frac{2}{3}(\hat{K}+2\Theta)^2 - 16 \pi \rho_{\rm ADM} \right. \nonumber \\
	& \left. \phantom{\frac{0}{0}}\;\; -2\kappa_1 (2+\kappa_2)\Theta \right] + \beta^i \partial_i \Theta \\
	\partial_t \tilde{\Gamma}^i &= \tilde{\gamma}^{jk} \partial_j \partial_k \beta^i + \frac{1}{3}\tilde{\gamma}^{ij}\partial_j \partial_k \beta^k - 2\tilde{A}^{ij} \partial_j \alpha + 2\alpha \left[ \tilde{\Gamma}^i_{jk} \tilde{A}^{jk} + 6\tilde{A}^{ij}\partial_j \phi \phantom{\frac{0}{0}}\right. \nonumber \\
	& \left. \;\;\;-\frac{1}{3} \tilde{\gamma}^{ij}\partial_j (2\hat{K}+\Theta) -\kappa_1(\tilde{\Gamma}^i - \tilde{\Gamma}_{\rm d}^i) - 8\pi\tilde{\gamma}^{ij}S_j\right] \nonumber \\
	&\;\;\;+ \frac{2}{3} \tilde{\Gamma}_{\rm d}^i \partial_j \beta^j - \tilde{\Gamma}_{\rm d}^j\partial_j \beta^i +\beta^j\partial_j \tilde{\Gamma}^i.
\label{Z4c}
\end{align}	
Here the Ricci tensor $R_{ij}$ of $\gamma_{ij}$ can be decomposed into a conformal and the Ricci tensor $\tilde{R}_{ij}$ associated with $\tilde{\gamma}_{ij}$.  Additionally, the constraints associated with $Z_i$ are absorbed into the definition of $\tilde{R}_{ij}$:
\begin{align}
	R_{ij} &= R^{\phi}_{ij} + \tilde{R}_{ij} \\
	R^{\phi}_{ij} &= -2 \tilde{D}_i \tilde{D}_j \phi - 2\tilde{\gamma}_{ij} \tilde{D}_k \tilde{D}^k \phi \nonumber \\
	& \;\;\;\; +4 \left( \tilde{D}_i \phi \tilde{D}_j \phi -\tilde{\gamma}_{ij} \tilde{D}_k \phi \tilde{D}^k \phi \right) \\
	\tilde{R}_{ij} &= -\frac{1}{2}\tilde{\gamma}^{lm}\partial_i \partial_j \tilde{\gamma}_{lm} + \tilde{\gamma}_{k(i}\partial_{j)}\tilde{\Gamma}^k + \tilde{\Gamma}_{\rm d}^k\tilde{\Gamma}_{(ij)k} \nonumber \\
	& \;\;\;\; +\tilde{\gamma}^{lm}\left(2\tilde{\Gamma}^k_{l(i}\tilde{\Gamma}_{j)km} + \tilde{\Gamma}^k_{im} \tilde{\Gamma}_{klj}\right)
\label{Z4cR}
\end{align}

This formulation also introduces the algebraic constraints:
\begin{equation}
	\det\tilde{\gamma} = 1, \;\;\;\;\; {\rm Tr}A \equiv \tilde{\gamma}^{ij}\tilde{A}_{ij} = 0.
\label{AlgbraicConstraint}
\end{equation}

Finally, to close the system of PDEs, we must introduce gauge fixing of the lapse $\alpha$, and shift $\beta$.  For most of the systems considered in this project we will implement the puncture gauge conditions composed of the Bona-Masso lapse \cite{Bona:1994dr} and the Gamma-Driver condition on the shift \cite{Alcubierre:2002kk}:
\begin{align}
	\partial_t \alpha &= -\mu_{\rm L} \alpha^2 \hat{K} + \beta^i \partial_i \alpha \\
	\partial_t \beta^i &= \mu_{\rm S} \alpha^2 \tilde{\Gamma}^i - \eta \beta^i + \beta^j \partial_j \beta^i
\label{gauge}
\end{align}
For most purposes the $1+\log$ variant is used with $\mu_L = 2/\alpha$ and $\mu_S = 1/\alpha^2$.

For completion of the Apples with Apples tests, we perform a gauge wave test below where we used the covariant conformal Z4 formulation (CCZ4) described in the appendix.  In this case we will also employ the harmonic lapse condition $\mu_L = 1$ with zero shift $\beta = 0$.

All of the dynamical tests considered in this project are performed in vacuum spacetimes with $\rho_{\rm ADM} = S_{ij} = 0$. 

\section{Numerical methods}

In this section we will describe in detail our numerical methods for evaluating the Z4 systems described in the previous section.  We intend to make our formulations and explanations as detailed as possible so other research efforts may compare with our results with no ambiguities in methods appearing.  This section will be devoted to the general numerical methods used for our simulations below.  Additional information specific to the particular problem will be described in the appropriate sections below.

\subsection{Discretization}
For spacetime solver applications we mostly employ the cell-centered data approach, however we occasionally make use of mixed nodal and cell centered data for certain one dimensional tests where determining convergence behavior was of primary interest.

For all problems considered, evolution is carried out using fourth order Runge-Kutta time integration.  The algebraic constraints (\ref{AlgbraicConstraint}) are enforced at each of the substeps of the Runge-Kutta time step. Spatial discretization is carried out using both second order and fourth order accurate finite differencing depending on the problem of interest.  For second order second order accurate problems centered discretization is performed as:
\begin{align}
	\partial_i \rightarrow D_{0i}, \;\;\;\;
	\partial_i \partial_j \rightarrow
	\begin{cases}
		D_{0i}D_{0j},& \text{if } i\neq j\\
		D_{+i}D_{-i}              & \text{if } i = j\\
	\end{cases},
\end{align}
where
\begin{align}
	&D_+v_j \equiv \frac{v_{j+1}-v_j}{\Delta x} \nonumber \\
	&D_-v_j \equiv \frac{v_j - v_{j-1}}{\Delta x} \nonumber \\
	&D_0 v_j \equiv \frac{v_{j+1}-v_{j-1}}{2\Delta x} \nonumber \\
	&D_+D_-v_j \equiv \frac{v_{j+1}-2v_j + v_{j-1}}{\Delta x^2}.
\end{align}
For fourth order accurate problems the spatial derivatives are defined for centered discretization as:
\begin{align}
	&\partial_i \rightarrow D_i^{(4)} \equiv D_{0i} \left(1-\frac{\Delta x^2}{6} D_{+i}D_{-i}\right) \nonumber \\
	&\partial_i \partial_j \rightarrow D_{0i}^{(4)} \equiv
	\begin{cases}
		D_{0i}^{(4)} D_{0j}^{(4)},& \text{if } i\neq j\\
		D_{+i}D_{-j}\left(1 - \frac{\Delta x^2}{12} D_{+i}D_{-i}\right)              & \text{if } i = j \\
	\end{cases}.
\end{align}

To accommodate the advection terms $\beta^i \partial_i f$ appearing in the Z4c system (\ref{Z4c}) and gauge conditions (\ref{gauge}), we make use of upwinded discretization:
\begin{align}
	\partial_i \rightarrow D_{i}^{\rm (up)} =
	\begin{cases}
		D_{+i} - \frac{1}{2} \Delta x_i D_{+i}D_{+i} & \text{if } \beta^i > 0\\
		D_{-i} + \frac{1}{2} \Delta x_i D_{-i}D_{-i} & \text{if } \beta^i < 0\\
	\end{cases}.
\end{align}

Finally, to stabilize high frequency modes which are excited in the numerical evolution, we employ Kreiss-Oliger dissipation whereby the right hand sides (RHS) of the equations of motion are modified:
\begin{equation}
	\partial_t \textbf{f} = {\rm RHS}(\textbf{f}) \rightarrow \partial_t \textbf{f} = {\rm RHS}(\textbf{f}) + Q \textbf{f},
\end{equation}
where the operator $Q$ is defined as $Q = Q_x + Q_y + Q_z$, with
\begin{equation}
	Q_x \equiv \sigma_{\rm KO} (-1)^{r+1} \Delta x^{2r-1}(D_+)^r (D_-)^r/2^{2r},
\end{equation}
and correspondingly for $Q_y$ and $Q_z$.  Here $r$ is defined as the $r \equiv \mathcal{O}/2 + 1$ for an order $\mathcal{O}$ accurate finite differencing scheme.  For our purposes we usually set $\sigma_{\rm KO} \sim 0.1$, however for CCZ4 problems a higher value $\sigma_{\rm KO} \sim 0.5$ is necessary. 

\subsection{AMR}

To support the multiple length scales within black hole binary simulations (evolution near the event horizon, and waveform extraction in the linear region), as well as ensuring that boundary conditions don't interfere with the physical region of interest, most spacetime evolution simulations rely on either fixed (FMR) or adaptive mesh refinement (AMR) techniques.  For this project we have built our code in the AMReX software framework.

\begin{figure}
	\centering
	\includegraphics[width=0.5\linewidth]{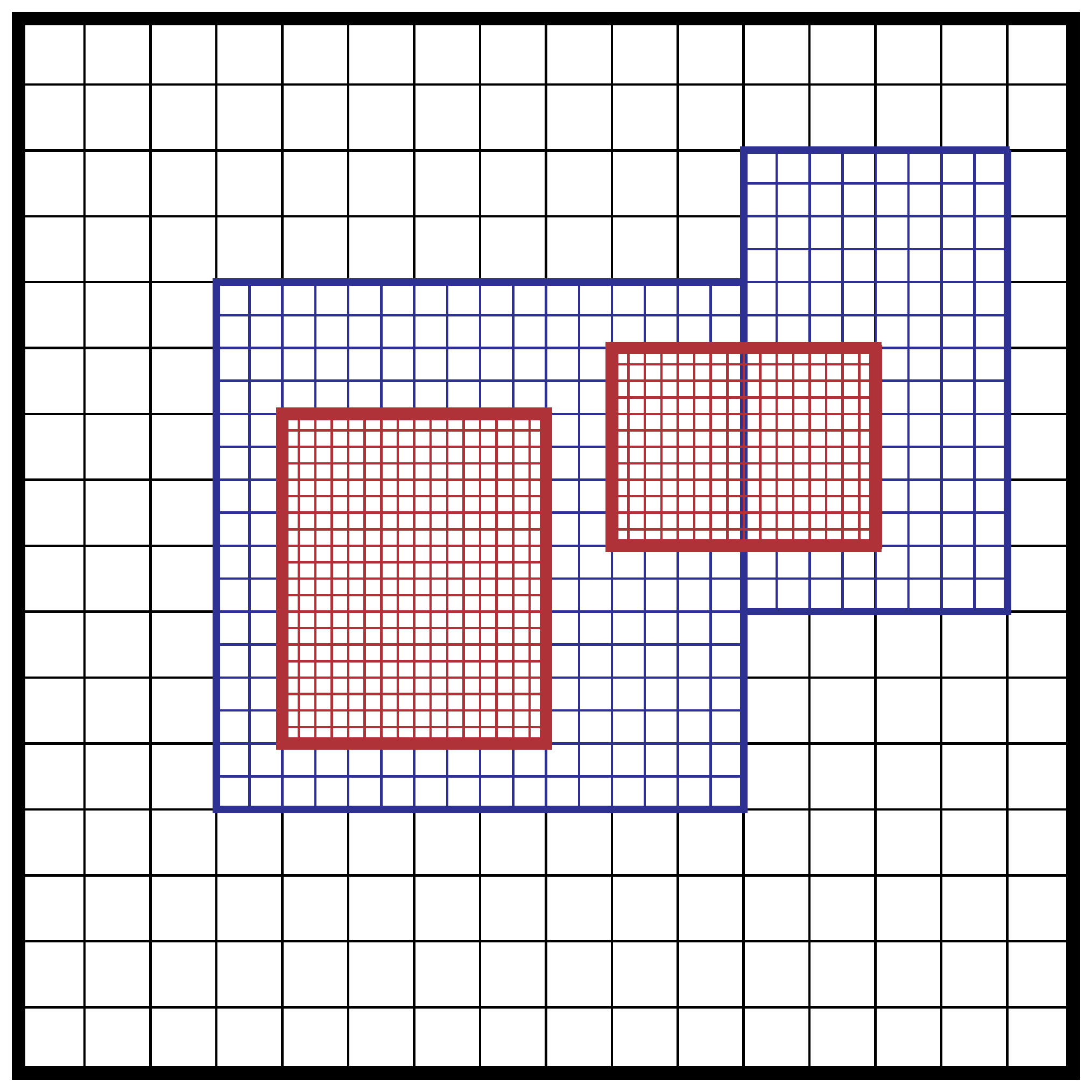}
	\caption{Illustration of AMR grids with two levels of factor 2 refinement.
		The coarsest grid covers the domain with 162 cells. Bold lines represent grid
		boundaries. The two intermediate resolution grids are at level 1 and the cells
		are a factor of two finer than those at level 0. The two finest grids are at level 2
		and the cells are a factor of 2 finer than the level 1 cells. Note that the level 2
		grids are properly nested within the union of level 1 grids, but there is no direct
		parent–child connection.}
	\label{AMRGridsIllustration}
\end{figure}

AMReX is a software framework for providing massively parallel, block-structured adaptive mesh refinement (AMR) applications.  AMReX uses a nested hierarchy of logically rectangular grids as illustrated in Figure \ref{AMRGridsIllustration}.  Here refined grid domains are always bounded by their parent grid domain, though they may intersect multiple blocks within that parent grid.  Each grid block is also formulated with appropriate ghost cells which are either filled with boundary conditions (if the block intersects the domain boundary) or are filled by interpolation from the next coarser grid.  The interpolation procedure is described in the following paragraphs.

\begin{figure}
	\centering
	\includegraphics[width=0.75\linewidth]{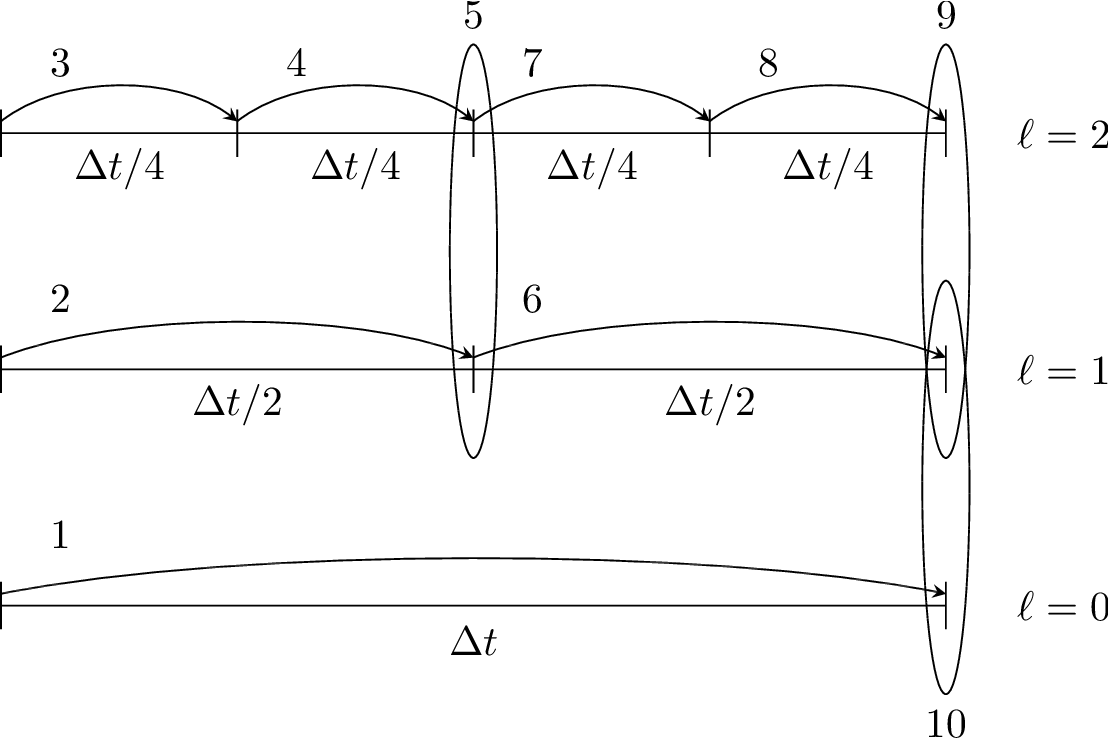}
	\caption{The subcycling approach is illustrated as a nested sequence if time substeps in each level.  Here the sequence of level advancement is shown for the example of 2 refinement levels (in addition to the coarse level) and can be extrapolated to any max level.  Higher levels are advanced with smaller time steps and performed in the order shown (labeled as steps 1 through 9).}
	\label{SubcyclingIllustration}
\end{figure}

Refinement of grids takes place simultaneously in both space and time.  Our particular AMReX applications use the subcycling-in-time approach, whereby finer grids are evolved multiple times with smaller time steps than the coarser levels.  Figure \ref{SubcyclingIllustration} illustrates the sequence of time step evolution at specific refinement levels.

For all FMR and AMR applications used in this project, during regridding, refined levels are filled using the cell conservative quartic method.  In this method a 4th order polynomial is used to fit the data.  For each cell involved in constructing the polynomial, the average of the polynomial inside that cell is equal to the cell averaged value of the original data on the coarser level.  We find that this approach (of the standard interpolations available in the AMReX toolkit) leads to the best stability and minimizes reflections of constraint violations at refinement boundaries. Coarse level data that is covered by a finer grid is updated by averaging down the fine level data after each RK stage.

To maintain 4th-order time and spatial accuracy at coarse-fine boundaries we perform a 4th-order accurate time interpolation of fine ghost cells at each stage of the Runge-Kutta procedure.  In this procedure approximations for fine level ghost data at each intermediate stage of the RK procedure are determined by interpolating solutions and derivatives at the coarse level using the RK stage values $k_1, ..., k_4$.  The specifics of this procedure are illustrated in great detail in section 3.1 of \cite{McCorquodale:2011}.

\subsection{Specifics of AMReX}

The AMReX framework includes several parameters for the user to select in order to optimize refinement and regridding, as well as optimizing memory distribution for parallel processes.  Throughout this project we include specifications of these parameters so future researchers may recreate our results with as little ambiguity as possible.  We describe the meaning and use of these additional parameters here.

During the gridding and regridding processes, at each level of refinement the domain is divided into logically rectangular blocks of cells based on the parameters specified at run time.  When dividing the region, cells in a block must be divisible by the blocking factor (in each direction, though we use one value for all directions in this project), and blocks may contain no more cells than specificed by the max grid size.

Finally, when tagging cells for refinement a box is grown in each direction by a minimum of the number of cells specified. Throughout this work we use one buffer cell when tagging and regridding.

\section{AMReX code generator}

The complexity of the Einstein equations presents a major hurdle towards writing efficient code in low level languages, where the barrier is significant between symbolic tensorial expressions and the executable blocks of code.  Many high level languages such as python and mathematica provide tensor manipulation packages that can be used to perform index manipulation within tensor calculations.

However, symbolic outputs from high level languages need significant manipulation to make equivalent low level executable blocks.  One example occurs when one wishes to write finite differenced equivalents of partial derivatives especially for higher order accuracy to fourth order and beyond.  Equally prevalent is the case of tensor index expansion.

The case of writing executable code blocks for Einstein's equations are particularly involved when one considers the requirement of fourth (or higher) order finite differencing with multiple centering schemes, and the complexity of the tensor expressions themselves.  Additionally, one is required to write right hand side expressions for upwards of 20 components, as well as perform post-updating corrections after each Runge-Kutta time step in the integration.

Many packages have been developed (such as Kranc and nrpy+) to overcome this hurdle, by automating the process of translating symbolic (textbook) expressions to executable blocks in low level languages such as C/C++ or Fortran.  One goal for this project was to develop an equivalent method of code generation to translate the symbolic Z4c expressions (\ref{Z4c}) into executable lines of code for AMReX, which itself is built on C++ and Fortran.  The AMReX Code Generator is built using Sympy to ease the manipulation and expansion of tensors and finite difference derivatives, as well as make use of general definitions of Christoffel symbols and curvature tensors from their formulations from the metric tensor.  Thus the user has to spend less effort expanding such definitions, and instead may rely on the code generator to produce the correctly expanded expressions.

As an example, if one is considering a simple Klein-Gordon system written in first order in time derivatives:

\begin{align}
	&\partial_t \psi = \pi \nonumber \\
	&\partial_t \pi = m^2 \psi + \nabla^2 \psi
\end{align}
In a python script one would write:

\begin{verbatim}
    psi = stvar(`psi', state = True)  \\Declaration of evolved variables
    pi = stvar(`pi', state = True)
		  
    dtpsi = stvar(`RHS_psi', rhs = True)  \\Declaration of right hand sides
    dtpi = stvar(`RHS_pi', rhs = True)
		  
    m = stvar(`m')  \\Constant mass term
		  
    dDDpsi_LL.expr = psi.diff('dDD', Accuracy = 2)  \\2nd order, 2nd Derivative
		  
    dtpsi.expr = pi.symb
    dtpi.expr = m**2*psi.symb**2 + dDDpsi_LL[0][0].symb + dDDpsi_LL[1][1].symb
\end{verbatim}
One may then output these variables in executable form, such as:
\begin{verbatim}
    print(psi.symb2isymb)
    print(dtpsi.symb2expr)
\end{verbatim}
with the resulting output:
\begin{verbatim}
    psi = state(i, j, k, Idx::psi); 
    rhs(i, j, k, Idx::psi) = std::pow(m,2)*psi + dDDpsi_LL_00 + dDDpsi_LL_11;
\end{verbatim}

Additionally, one may output appropriately expanded finite differencing formulas for numerical derivatives to arbitrary accuracy and with offsetting for up/down-winding.  For example, the 4th order central differenced first derivatives of $\psi$ can be accessed using:

\begin{verbatim}
    print(dDpsi_L.symb2expr())
\end{verbatim}
with output
\begin{verbatim}
    amrex::Real dDpsi_L_0 = ((2.0/3.0)*state_fab(i + 1, j, k, Idx::psi) 
                            - 1.0/12.0*state_fab(i + 2, j, k, Idx::psi) 
                            - 2.0/3.0*state_fab(i - 1, j, k, Idx::psi) 
                            + (1.0/12.0)*state_fab(i - 2, j, k, Idx::psi))/dx[0];
    amrex::Real dDpsi_L_1 = ((2.0/3.0)*state_fab(i, j + 1, k, Idx::psi) 
                            - 1.0/12.0*state_fab(i, j + 2, k, Idx::psi) 
                            - 2.0/3.0*state_fab(i, j - 1, k, Idx::psi) 
                            + (1.0/12.0)*state_fab(i, j - 2, k, Idx::psi))/dx[1];
\end{verbatim}
In this case, the domain dimension has been defined as ${\rm dim} = 2$, but up to ${\rm dim} = 3$ is supported.

\section{AwA tests and results}

In order to validate the spacetime solver built by the code generator discussed in the previous section, we perform a selected set of tests from the standard Apples with Apples testbeds for Einstein solvers.  To this we perform tests designed to validate the accuracy and convergence of the solvers under refinements of the numerical grid.  

We include the standard robust stability test to probe the constraint damping behavior in both 2nd and 4th order finite differencing.  We also perform one dimensional linear wave (2nd order finite differencing) and gauge wave tests (4th order finite differencing) for both convergence testing and stability for long time evolutions.  These tests are also useful in that they require no numerically constructed initial data, and instead involve exact initial data.  

Finally, we perform black hole binary simulations for both head on collisions and binary inspirals for equal mass black holes.  Head on collisions also involve exact initial data via the puncture approach.  They also provide a straightforward method for testing convergence behavior in waveform extractions.  Binary inspirals are more difficult to asses for accuracy, however it is useful to compare both the waveforms and black hole trajectories with other previously constructed spacetime solvers.

\subsection{Robust stability test}

The Robust stability tests serves to diagnose the constrain violation behavior of a particular method of solving Einstein's equations.  It has most notably served to characterize stability behavior for solvers based on the $3+1$ ADM, BSSN, Z4, etc.  It is well known that the ADM solvers suffer from rapid undesirable growth of constraint violation that effectively renders solver based on this method useless.  It has also been demonstrated that the BSSN methods develop moderate but bounded growth of constraint violations, and the Z4c methods result in decreasing constraint violations.

Here we simply wish to validate our solver based on code generation in the Z4c scheme with puncture gauge conditions in both 2nd order and 4th order finite differencing. To this end we expect to observe decreasing constraint violation behavior.

For this test we perform the simulations on a cell centered grid of $n_x = 64\rho$ points in the x direction.  For 2nd order tests we use $n_y = n_z = 4\rho$, and for 4th order tests $n_y = n_z = 8\rho$. We perform the test on a domain of $x \in (-0.64, 0.64)$.  For 2nd order tests $y$ and $z \in (-0.04,0.04)$.  For 4th order we have  $y$ and $z \in (-0.08,0.08)$.  All 22 variables are initialized randomly $\varepsilon = (-10^{-10}/\rho^2,10^{-10}/\rho^2)$.  Evolution is performed with $\eta = 2$, $\kappa_1 = \kappa_2 = 0$, $\sigma_{\rm KO} = 0.1$.  We perform the test for a CFL factor of 0.5 for 1000 light crossing times ($t_{\rm final} = 1280$) on the domain.  

The constraint violations are monitored using the suggested diagnostic variable \cite{Cao:2011fu}:

\begin{equation}
	C \equiv \sqrt{H^2 + \gamma_{ij}M^i M^j + \Theta^2 + 4\gamma_{ij}Z^i Z^j},
\label{ConstraintMonitor}
\end{equation}
where $H$ and $M^i$ are the Hamiltonian and Momentum constraints defined in (\ref{PhysicalConstraints}).

\begin{figure}
	\centering
	\includegraphics[width=0.75\linewidth]{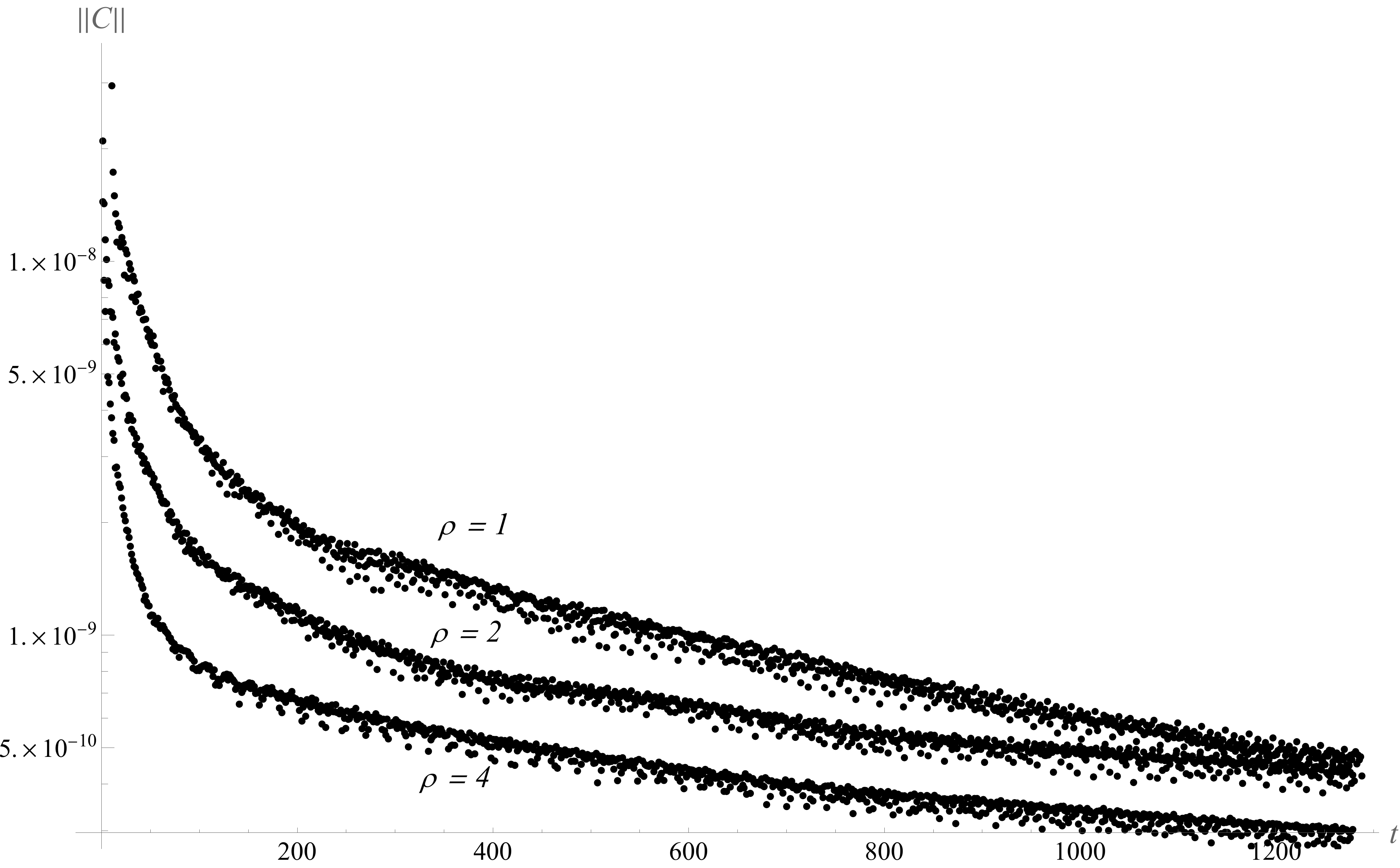}
	\caption{Here the norm of the constraint monitor $||C||$ is graphed for the robust stability test.  Second order finite differencing is used for spatial derivatives in the evolution equations.  Three values for $\rho$ were considered for amplitudes $\varepsilon = (-10^{-10}/\rho^2,10^{-10}/\rho^2)$.}
	\label{RobustStability2ndOrder}
\end{figure}

\begin{figure}
	\centering
	\includegraphics[width=0.75\linewidth]{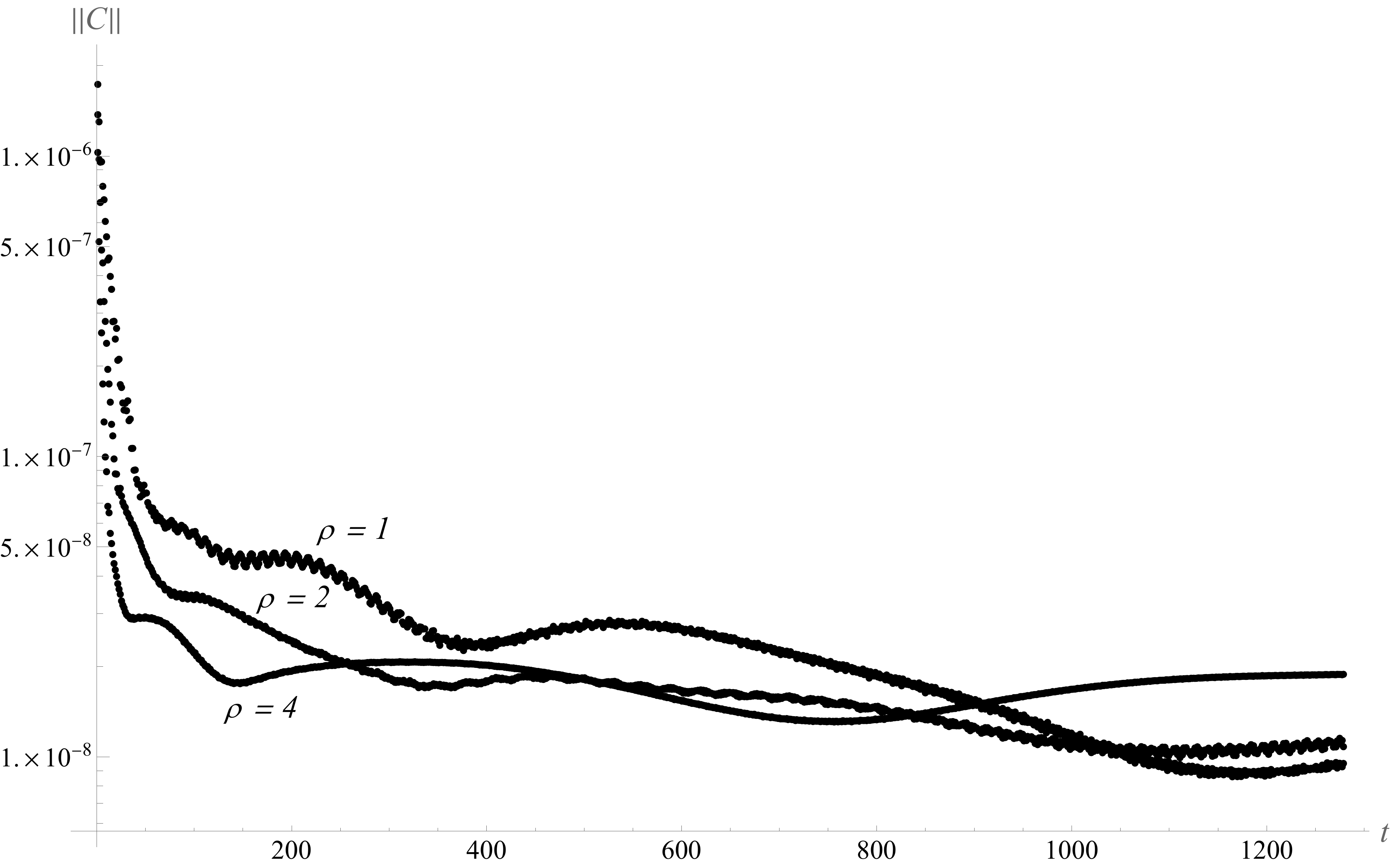}
	\caption{Here the norm of the constraint monitor $||C||$ is graphed for the robust stability test.  Fourth order finite differencing is used for spatial derivatives in the evolution equations.  Three values for $\rho$ were considered for amplitudes $\varepsilon = (-10^{-8}/\rho^2,10^{-8}/\rho^2)$.}
	\label{RobustStability4thOrder}
\end{figure}

The results for both 2nd and 4th order finite differencing are shown in figures (\ref{RobustStability2ndOrder}) and (\ref{RobustStability4thOrder}) respectively.  In each case we observe the expected constraint damping behavior of the diagnostic variable $C$.

\subsection{Linear wave test}

We test numerical stability and precision of our generated solver by evolving a linear wave as a perturbation on the Minkowski background.  In this case a linear wave is an exact solution of the linearized Einstein equations which are valid for small amplitudes where interaction terms are below threshold for generating transitions between modes.  Thus a linear wave of sufficiently small amplitude should maintain its structure up to numerical accuracy and machine precision.

To this end we use analytic initial data of the form:
\begin{align}
	\tilde{\gamma}_{xx} &= 1, & 	\gamma_{yy} &= 1+b, & \gamma_{zz} &=1-b  \\
	\alpha &= 1, & 	K_{yy} &= \frac{1}{2}\partial_t b, & 	K_{zz} &= -\frac{1}{2}\partial_t b,
\end{align}
with 
\begin{equation}
	b = A \sin\left(\frac{2\pi(x-t)}{d}\right),
\label{bwave}
\end{equation}
where $A = 10^{-8}$ and $d = 1.28$.  The amplitude $A$ is chosen such that non-linear terms in (\ref{Z4c}) are below machine precision and can thus be numerically neglected in the evolution.

The test is performed for 2nd order on a (node, cell, cell) centered grid with $n_x = 64\rho$, $n_y = n_z = 4\rho$ on the domain $x \in (-0.64,0.64)$, and $y,z \in (-0.04,0.04)$.  We perform the test for $\rho = 1,2,4,$ and $8$.  Here we choose  $\eta = 2$, $\kappa_1 = 0.02$ $\kappa_2 = 0$, with dissipation factor $\sigma_{\rm KO} = 0.1$.  We chose the Courant factor as $\lambda = 0.5$.

\begin{figure}
	\centering
	\includegraphics[width=0.75\linewidth]{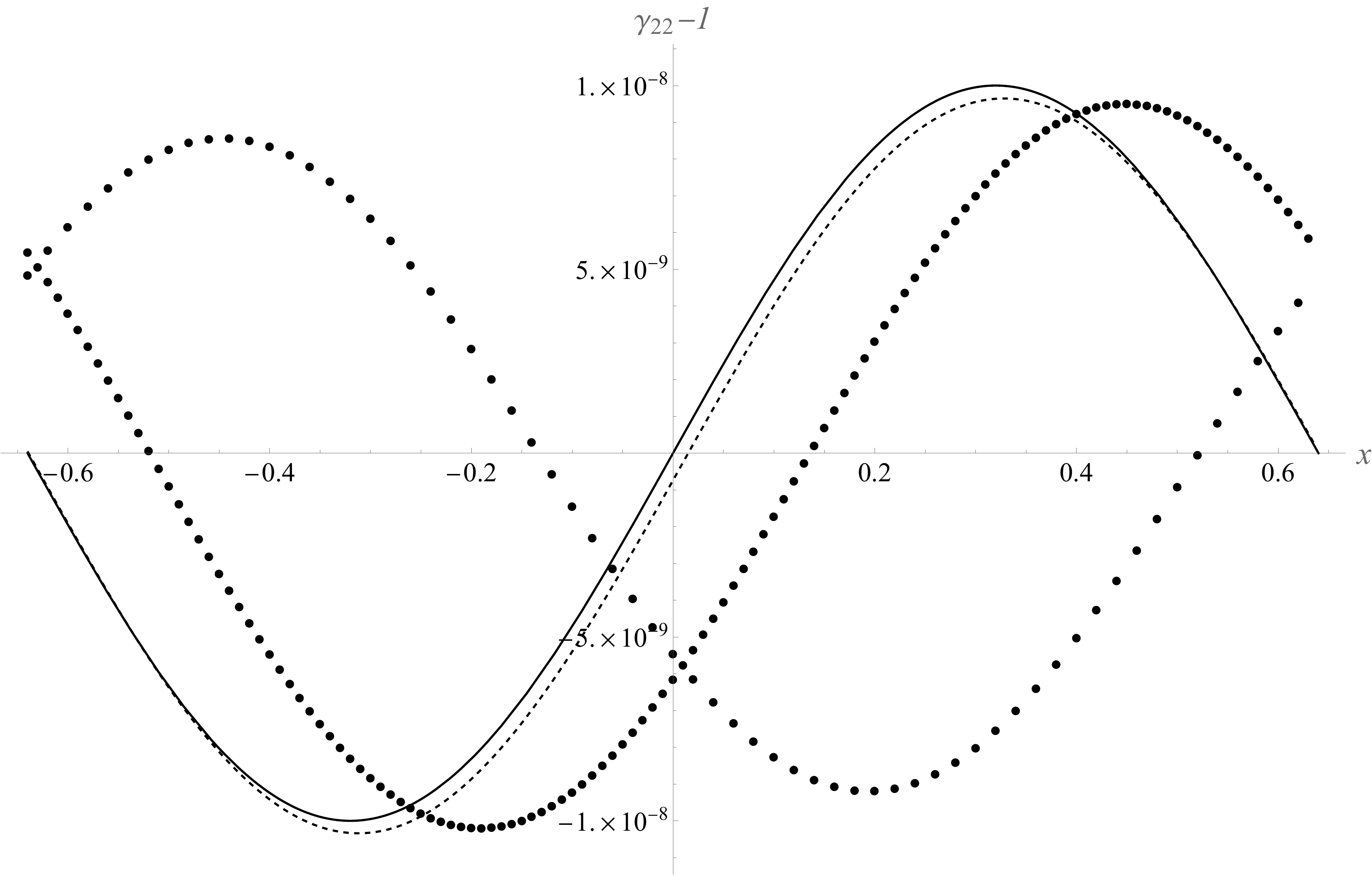}
	\caption{Comparison of wave forms for linear wave initial data at $t = 1000$ crossing times.  Here we compare the wave forms for $\gamma_{11} - 1$. Resulting forms for $\rho = 1,2$ are shown (sparsely dotted, and dotted respectively).  The resulting form for $\rho = 8$ is also shown (dashed line), along with the exact solution (solid line).}
	\label{LinearWaves}
\end{figure}

\begin{figure}
	\centering
	\includegraphics[width=0.75\linewidth]{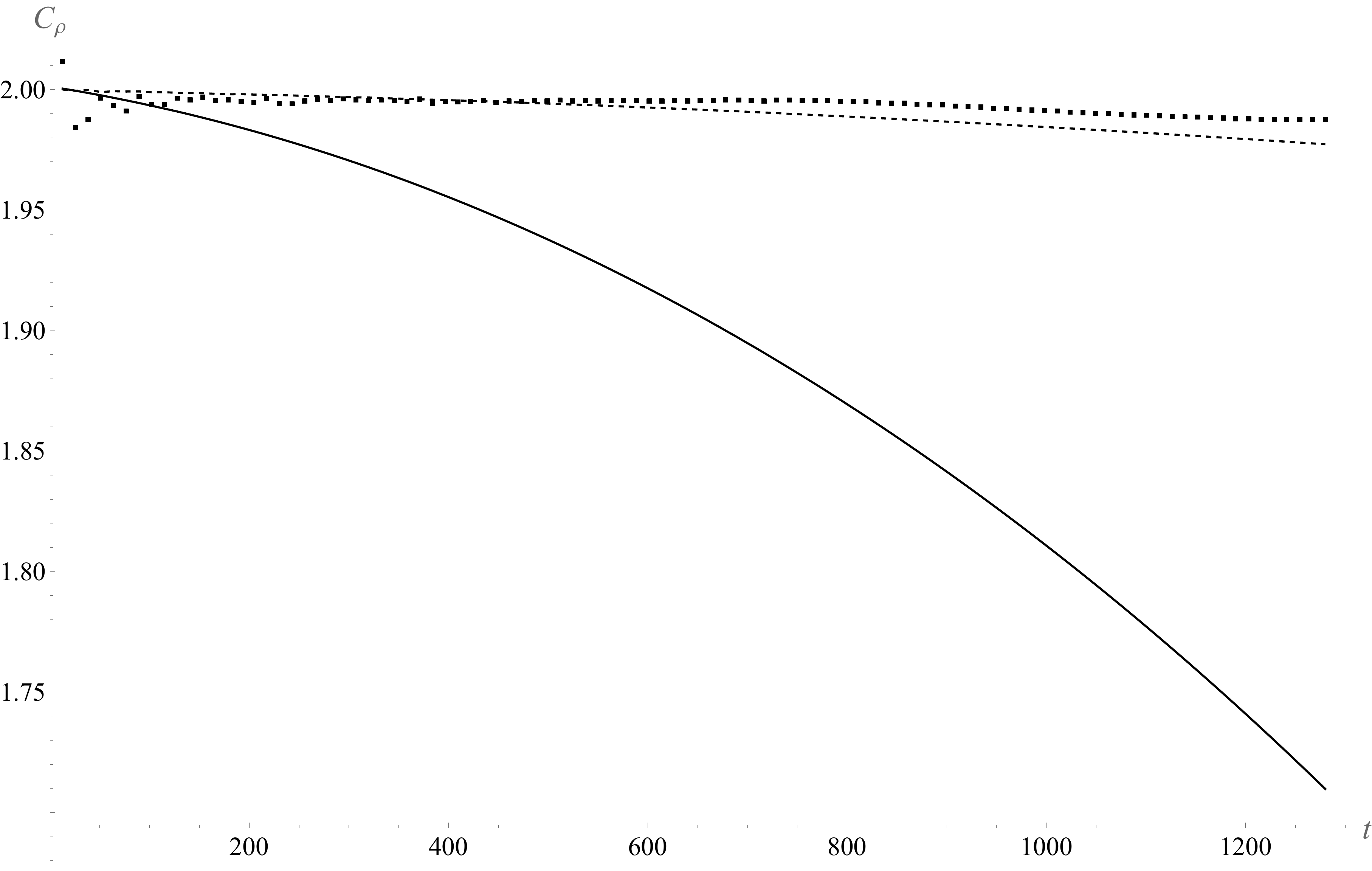}
	\caption{Convergence of linear wave forms for 2nd order finite differencing using the convergence formula in (\ref{ConvergenceFormula}). Here we compare waveforms for $\rho = 1,2,4,8$, and $16$ using the formulas for $C_{\rho = 4}$ (solid), $ C_{\rho = 8}$ (dashed), and $C_{\rho = 16}$ (dotted).}
	\label{LinearWavesLogConvergence}
\end{figure}

Figure \ref{LinearWaves} illustrates the $\gamma_{yy}$ wave forms for a linear wave at $t = 1000$ crossing times.  The convergence of solutions is apparent in these plots.  The convergence of increasing resolutions is further illustrated in Figure \ref{LinearWavesLogConvergence}.  Here the convergence is defined by comparing results at three different resolutions via the formula:
\begin{equation}
    C_\rho \equiv \frac{||u_{\rho} - u_{\rho/2}||}{||u_{\rho/2} - u_{\rho/4}||},
    \label{ConvergenceFormula}
\end{equation}
where the norm $||u||$ is defined as the sum of squares of the 22 evolved variables in (\ref{Z4c}) and (\ref{gauge}):
\begin{equation}
    ||u||^2 = \sum_i u_i^2 = \phi^2 + \tilde{\gamma}_{ij}^2 + \hat{K}^2 + \tilde{A}_{ij} + {\rm ...}
\end{equation}
We will use this definition for both the linear wave and gauge wave tests below (with appropriate variables redefined when the CCZ4 formulation is used.

\subsection{Gauge wave test}

As a final quantitative test for our code generated solvers, we employ the gauge wave test witch 4th order finite differencing.  Gauge wave tests serve to illustrate several key characteristics of convergence of Einstein solvers.  In particular, gauge wave tests allow one to probe the numerical stability and accuracy of the solver in a technically non-perturbative manor (arbitrary wave amplitude), by introducing an exact time dependent solution of Einstein's equations that are gauge equivalent to vacuum spacetime.  Thus one is able to illustrate how well the solver maintains gauge invariance.  Typically this is achieved by monitoring the Hamiltonian and momentum constraints (\ref{PhysicalConstraints}) for long time evolutions.

For this test we initialize the variables according to an exact gauge wave:
\begin{align}
	\tilde{\gamma}_{xx} &= 1-b, & 	\gamma_{yy} &= 1, & \gamma_{zz} &=1  \\
	\alpha &= \sqrt{1-b}, & 	K_{xx} &= \frac{\partial_t b}{2\sqrt{1+b}}, & \hat{\Gamma}^x &= -\frac{2 \partial_x b}{3(1-b)^{5/3}},
\end{align}
where $b$ is defined as in (\ref{bwave}) with $d = 1$.  Here we choose an amplitude $A = 0.1$.  Such an amplitude allows for non-linear terms in the equations of motion to affect the solution, and also allows for testing at 4th order finite differencing.  

For this test we again use a (node, cell, cell)-centered grid, with $n_x = 64\rho$, $n_y = n_z = 4\rho$ on the domain $x \in (-0.5,0.5)$, and $y,z \in (-0.03125,0.03125)$.  However, we choose to evolve the system in a conformal covariant Z4 (CCZ4) scheme, with harmonic lapse condition ($\mu_L = 1$), and zero shift.  Additionally, we set $\eta = 2$, $\kappa_1 = 1$, $\kappa_2 = 0$, and $ \kappa_3 = 1$, with dissipation factor $\sigma_{\rm KO} = 0.3$.  We evolve with a CFL factor of $\lambda = 0.5$.

\begin{figure}
	\centering
	\includegraphics[width=0.75\linewidth]{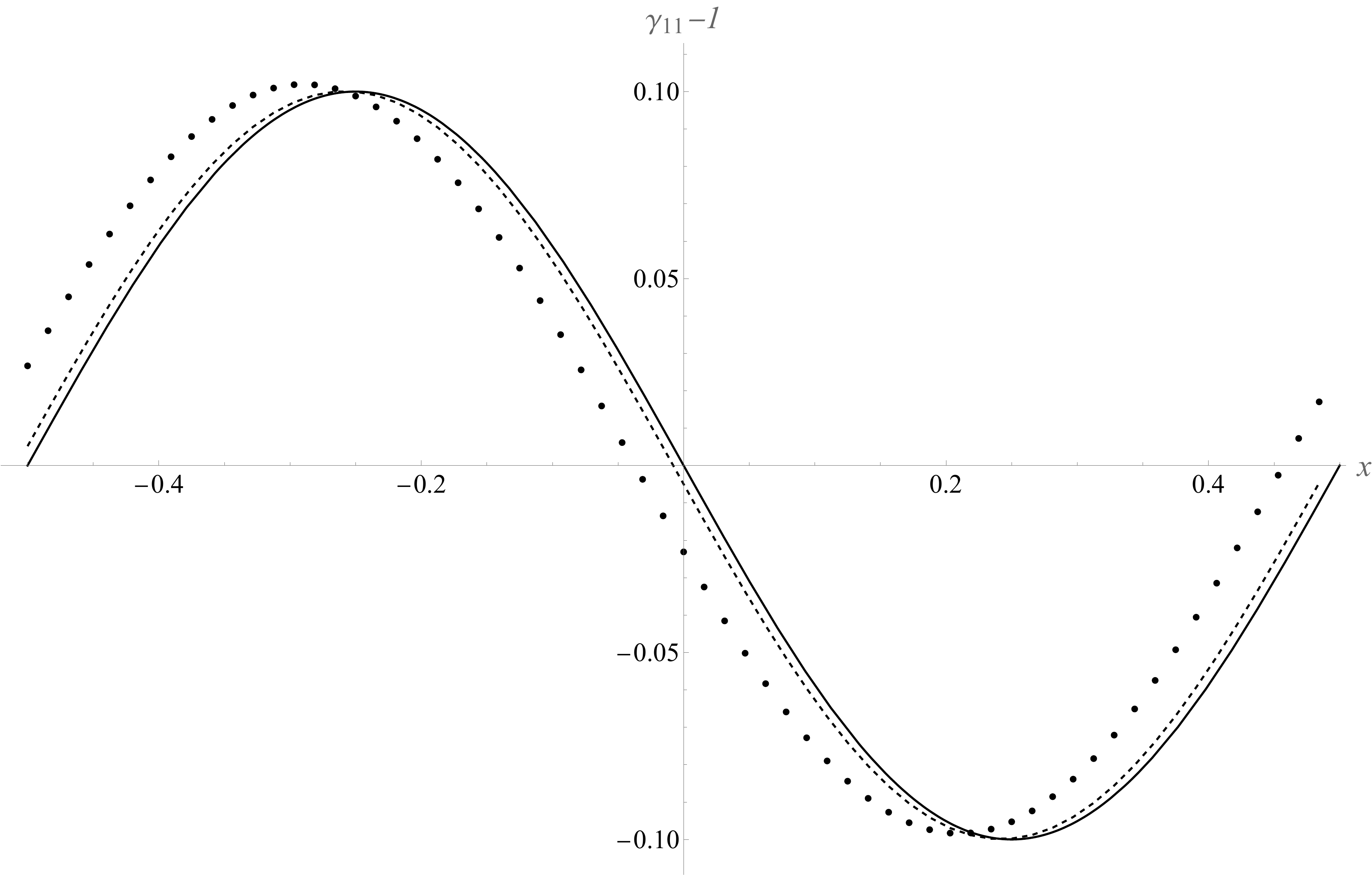}
	\caption{Comparison of gauge waves at $t = 1000$ crossing times, for $\rho = 4$ (dotted) and $\rho = 8$ (dashed).  Here we compare $\gamma_{11} - 1$. The exact solutions is plotted (solid line). Evolution is done in CCZ4 formulation with harmonic lapse and zero shift.}
	\label{GaugeWaves}
\end{figure}

\begin{figure}
	\centering
	\includegraphics[width=0.75\linewidth]{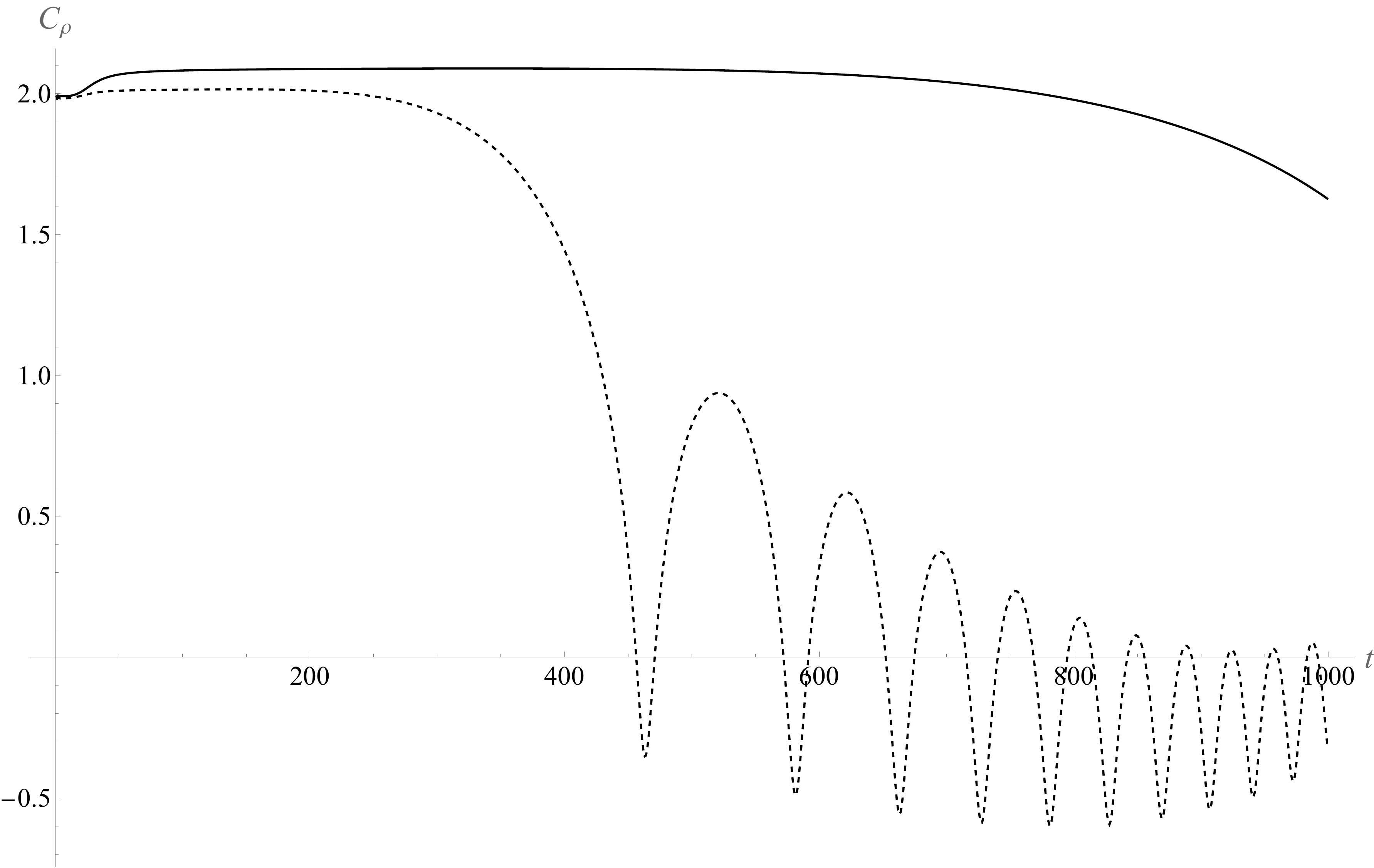}
	\caption{Convergence of waveforms for gauge waves with 4th order finite differencing.  Comparisons are performed for $\rho = 8$, $\rho = 4$, $\rho = 2$, and $\rho = 1$, using the formulas for $C_{\rho = 4}$ (dashed) and $C_{\rho = 8}$ (solid) given in (\ref{ConvergenceFormula}).}
	\label{GaugeConvergence}
\end{figure}

\begin{figure}
	\centering
	\includegraphics[width=0.75\linewidth]{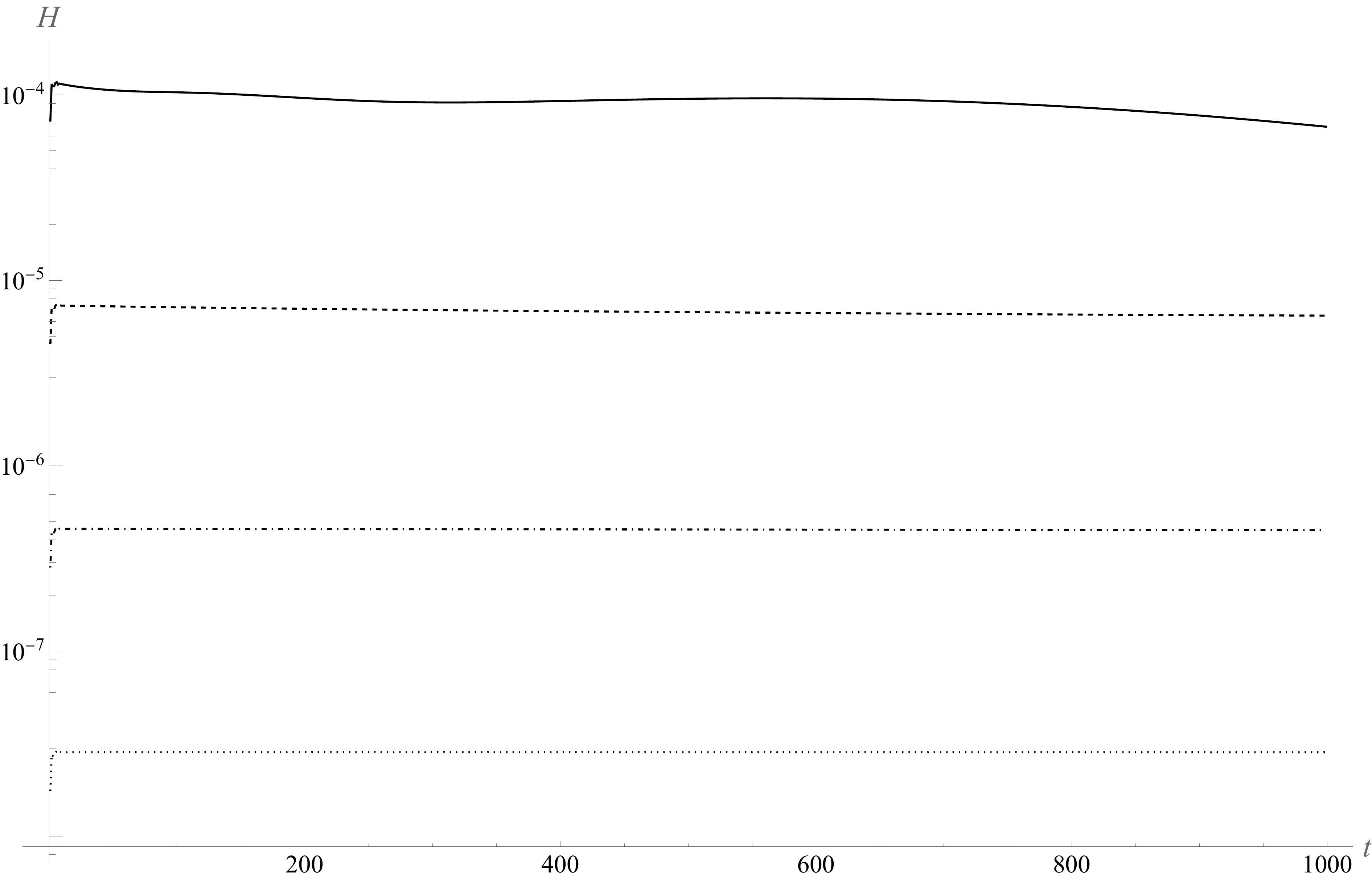}
	\caption{Shown is the $H$ constraint (\ref{PhysicalConstraints}) as a function of crossing times for $\rho = 1,2,4,$ and $8$ (solid, dashed, dot-dashed, and dotted respectively) for gauge waves with 4th order finite differencing.}
	\label{GaugeWaveH}
\end{figure}

\begin{figure}
	\centering
	\includegraphics[width=0.75\linewidth]{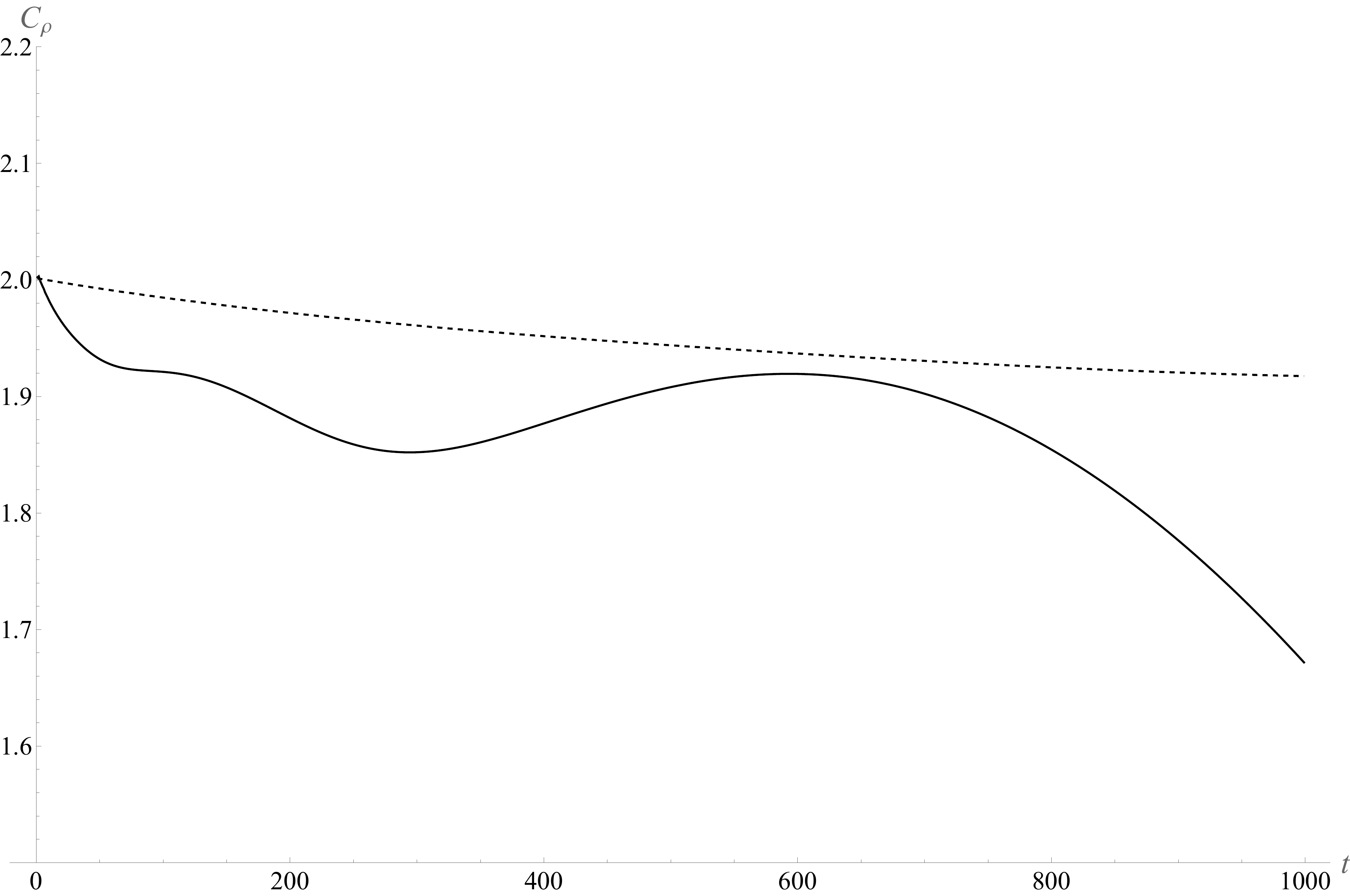}
	\caption{Convergence of $H$ constraint for gauge waves with 4th order finite differencing.  Here the convergence of $H$ is performed with (\ref{ConvergenceFormula}) using the single variable $u \equiv H$ for $C_{\rho = 4}$ (solid) and $C_{\rho = 8}$ (dashed).}
	\label{GaugeWaveHConvergence}
\end{figure}

Figure \ref{GaugeWaves} illustrates the convergence of wave forms at $t=1000$ crossing times for increasing $\rho$ to the exact solution.  The waveform convergence as a function of time is shown quantitatively in figure \ref{GaugeConvergence}.  Finally, we also monitor the Hamiltonian constraint \ref{PhysicalConstraints} as a function of time for all value of $\rho$ considered.  Shown in figure \ref{GaugeWaveH}, is the integrated value of $H$ as a function of time.  The convergence of $H$ at 4th order is shown in figure \ref{GaugeWaveHConvergence}.

\section{Black hole binary systems}

As a final set of tests for the code generated spacetime solver, we wish to demonstrate numerical simulations of black hole binary mergers.  Such tests serve both as interesting physical systems, as well as a strong test of numerical stability for 3 spatial dimensions.  We are also able to observe the behavior of physical simulations in the presence of FMR and AMR.

To perform these tests we will consider both a head on binary collision, as well as a binary inspiral.  In both cases we observe the gravitational radiation via the Weyl scalar $\psi_4$ in the Newman-Penrose tetrad formulation for waveform extraction.

The initial data for black hole binaries is generated in a separate solver from the evolution system (\ref{Z4c}).  

\subsection{Head on collision}

The goal of the equal mass black hole head on collision test is to quantitatively demonstrate the convergence behavior of the extracted wave forms for several resolutions.  Here multiple resolutions can be tested easily since the solution of the initial data \cite{Brill:1963yv} \cite{Brandt:1997xd} \cite{Okawa:2013afa} is exact for puncture black holes initially at rest on the initial spatial hypersurface.  In particular, it is clear from the Appendix that an exact solution of the constraints is achieved by simply setting $A_{ij} = K = 0$ and $u = 0$ in (\ref{psidecomposition}).  Thus, initial data for head on collisions is analytic, and no comparison of numerical initial data is necessary.  

For the initial data the equation (\ref{HamiltonianConstraintSimpleU}) is solved analytically with $u = 0$, and the definition of $\xi$ given in (\ref{xidefinition}).  Here we place two equal mass black hole punctures at rest with bare masses $M_\pm = 0.5$ at coordinate positions $\vec{x}_\pm = (0, \pm 1.1515, 0)$.  

In this case we use a cell centered numerical grid with $(n_x,n_y,n_z) = (64\rho, 64\rho, 64\rho)$ for $\rho = 1,2,4,$ and $8$ at the coarsest level, on th domain $x,y,z \in (-128, 128)$.  Here the spatial derivatives are approximated using 4th order finite differencing, however it should be noted that convergence will only be available to 2nd order as explained below.  We perform the evolution with 6 levels of FMR (with level $l = 0$ as the coarsest level) with a refinement factor of 2 at each level.  Cells are tagged for refinement with the criterion $r < 64,32,16,8,4,2$, for $r = (x^2 + y^2 + z^2)^{1/2}$ the standard radius on a Cartesian grid.  We choose a blocking factor of 16 for $\rho = 1,2$ and 32 for $\rho = 4,8$.  The max grid size is set to 32 cells.  One buffer cell is used. Regridding takes place every 10 timesteps.  The systems is evolved with the Z4c scheme, with puncture gauge conditioning.  For this case we set $\eta = 2$, $\kappa_1 = 0.02$, and $\kappa_2 = 0$.  We choose a dissipation factor of $\sigma_{\rm KO} = 0.1$.  The Courant factor is chosen to be $\lambda = 0.5$.

To analyze the simulation we perform a waveform extraction at various coordinate radii in the near linear regime $r = 20,30,40,50$.  The wave form is constructed using the Newman-Penrose formalism.  Following the construction of the Weyl scalar $\psi_4$, we determine the $l=2,$  $m = 0$ spherical harmonic amplitude $\psi_4^{20}$.  Since the data is native to Cartesian coordinates, we perform a trilinear interpolation to the sphere at the radius of extraction.  

\begin{figure}
	\centering
	\includegraphics[width=0.75\linewidth]{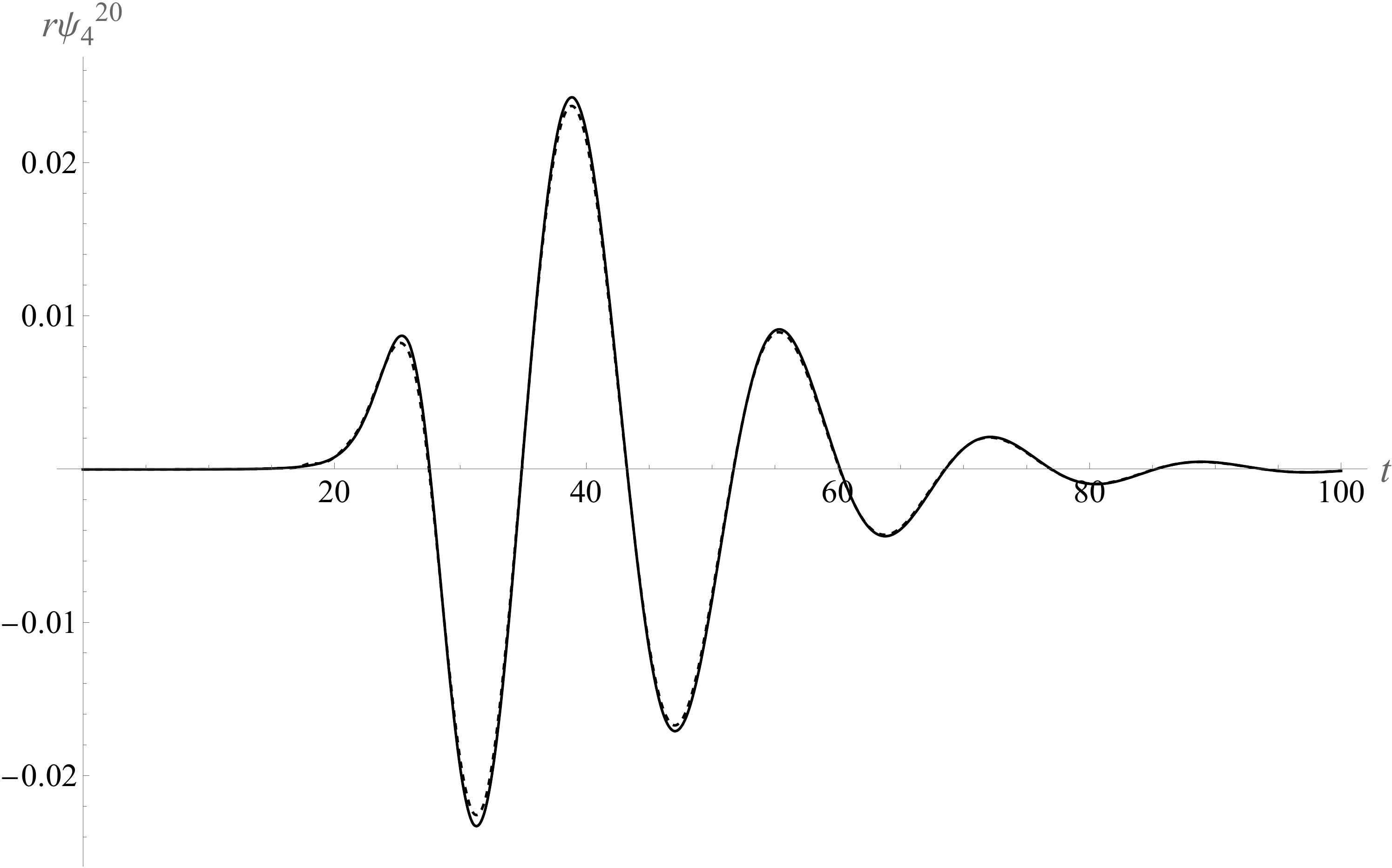}
	\caption{The $\psi_4^{20}$ waveform for the head-on black hole collision at $r = 20$ is shown for both the lowest resolution $n = 64$ (dashed) and the highest resolution $n = 256$ (solid).}
	\label{BHWaveFormFull}
\end{figure}

\begin{figure}
	\centering
	\includegraphics[width=0.75\linewidth]{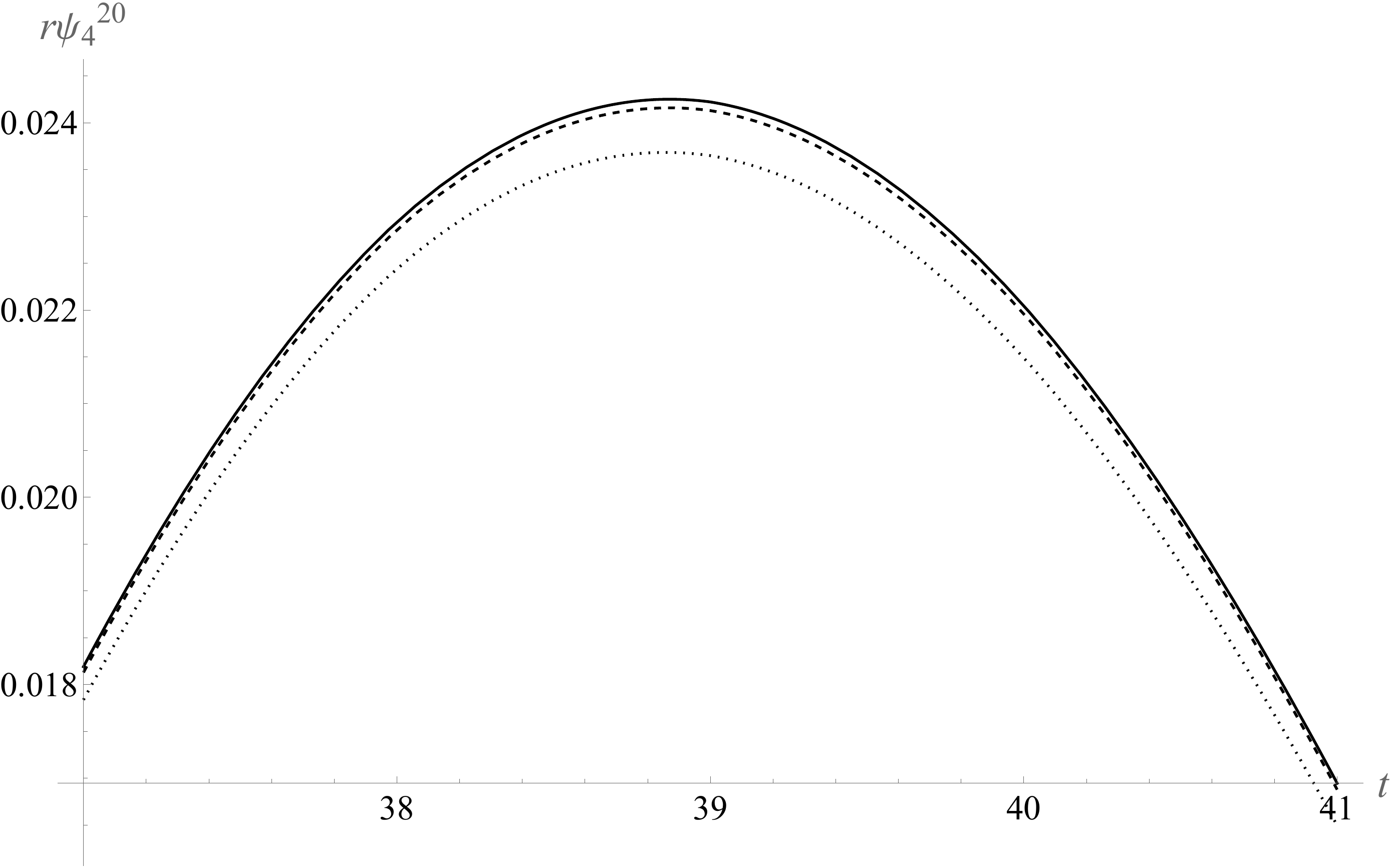}
	\caption{To illustrate the convergence of the waveform for a head on collision, we zoom in on the highest peak in figure \ref{BHWaveFormFull}.  All three resolutions are shown; lowest (dotted), medium (dashed), and highest (solid).}
	\label{BHWaveFormPart}
\end{figure}

As stated previously, evolution is performed using 4th order spatial finite differencing.  However, it can be shown that numerical interpolation and integration of cell centered Cartesian data on a spherical shell leads to an expected 2nd order convergence of the waveform for increasing resolution.  This is indeed observed as shown in figures \ref{BHWaveFormFull} and \ref{BHWaveFormPart}.

\begin{figure}
	\centering
	\includegraphics[width=0.75\linewidth]{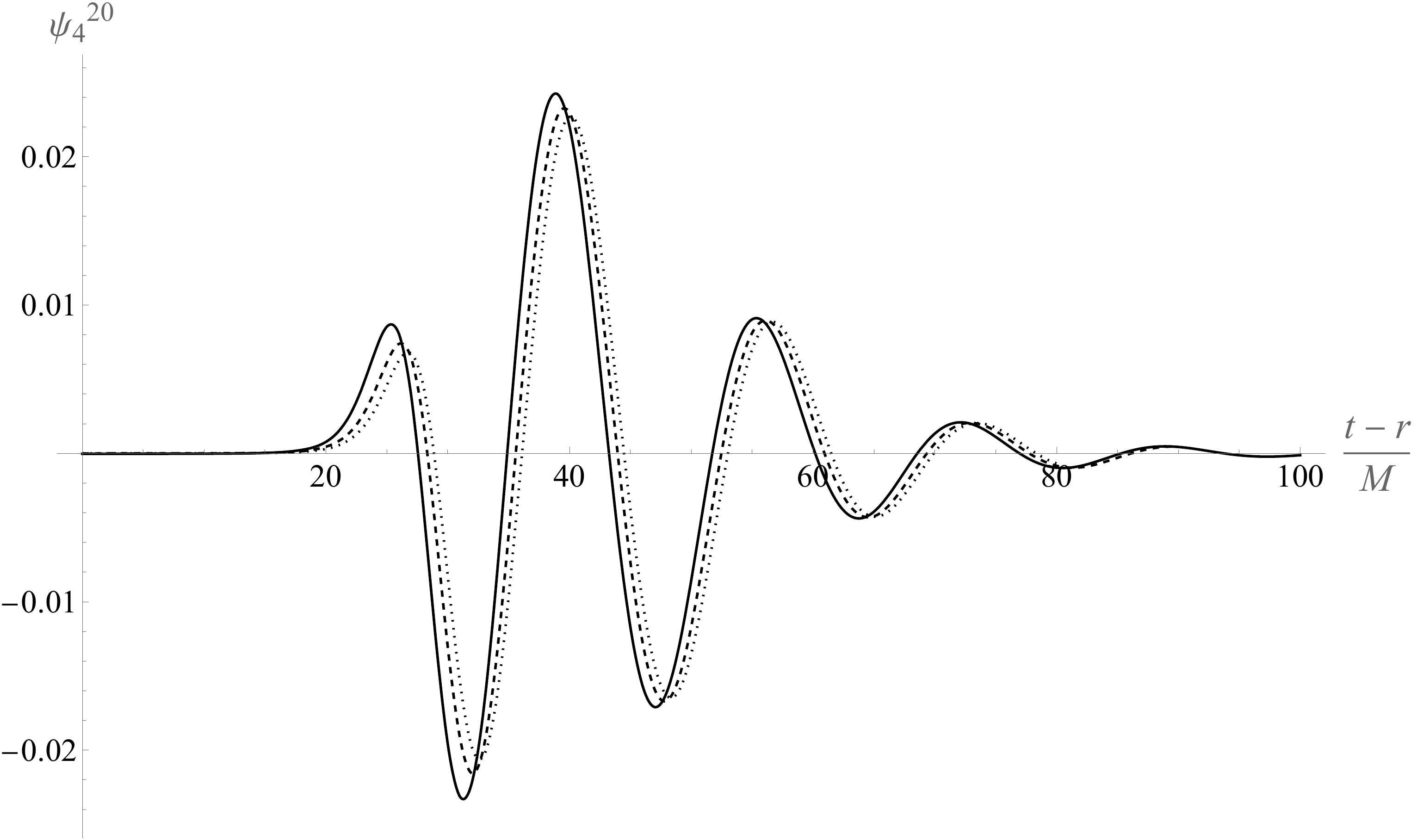}
	\caption{The robustness of the waveform is illustrated by plotting the waveform at 3 radii $r = 20$ (solid), $r = 30$ (dashed), and $r = 40$ (dotted).  The $r = 30$ and $r = 40$ waveforms are plotted with a $\Delta t = 10$ and $\Delta t = 20$ delay respectively and overlaid to illustrate the evolution of the waveform.}
	\label{}
\end{figure}

\subsection{Black hole inspiral}

The final test we wish to present is that of an equal mass black hole binary inspiral as one might expect to compare with observational data from gravitational wave interferometer observatories.  This last test has the added requirement of numerically generated initial data, that is fed into the Z4c solver, which will then serve as a test of both solvers.

For the initial data we solve (\ref{HamiltonianConstraintSimpleU}) using the solution for the momentum constraint (\ref{MomentumConstraintSolution}), and the definition of $\xi$ given in (\ref{xidefinition}).  Here we place two equal mass black hole punctures with bare masses $M_\pm = 0.4856$ at coordinate positions $\vec{x}_\pm = (0, \pm 4.891, 0)$.  Here the punctures are given initial momenta $\vec{P}_\pm = (\mp 0.0969, 0, 0)$.  The initial data system is solved using relaxation until the Hamiltonian constraint is satisfied to the order $H < \mathcal{O}(10^{-6})$ outside the black hole horizons.

Following the initial data solving procedure, we initialize the gauge variables as $\alpha = \psi^{-2}$, $\beta^i = 0$ as in Brill-Lindquist initial data.  Additionally, we initially set $\phi = \log{\psi}$ per the definition, and $\tilde{\gamma}_{ij} = \delta_{ij}$.  Additionally, we set $\hat{K} = \theta = \tilde{\Gamma}^i = 0$.

For this case we performed the simulation on a coarse grid of $(n_x,n_y,n_z) = (256, 256, 256)$, for a domain $x,y,z \in (-512, 512)$  with 4th order spatial finite differencing. In this case we use a mixed FMR/AMR grid with 8 levels of refinement, and a refinement factor of 2.  The first 4 levels are refined on a fixed grid as in the head on collision case with $r < 256,128,64,32$.  The last 4 levels are tagged using AMR on the lapse $\alpha < 0.8, 0.7, 0.6, 0.5$.  This type of mixed FMR/AMR allows for consistent interpolation and integration of waveforms to spherical shells at the radius of the extraction for some $r > 32$.  Here the blocking factor is chosen to be 8 with a max grid size of 16 with one buffer cell.  The AMR grid is updated after every 5 time steps in the RK4 evolution.  We evolve the equations of motion in the Z4c formalism in the puncture gauge with$\eta = 0.25$, $\kappa_1 = 0.02$, $\kappa_2 = 0$, and a dissipation factor  $\sigma_{\rm KO} = 0.1$.  We choose a Courant factor of $\lambda = 0.1$.

\begin{figure}
	\centering
	\includegraphics[width=0.5\linewidth]{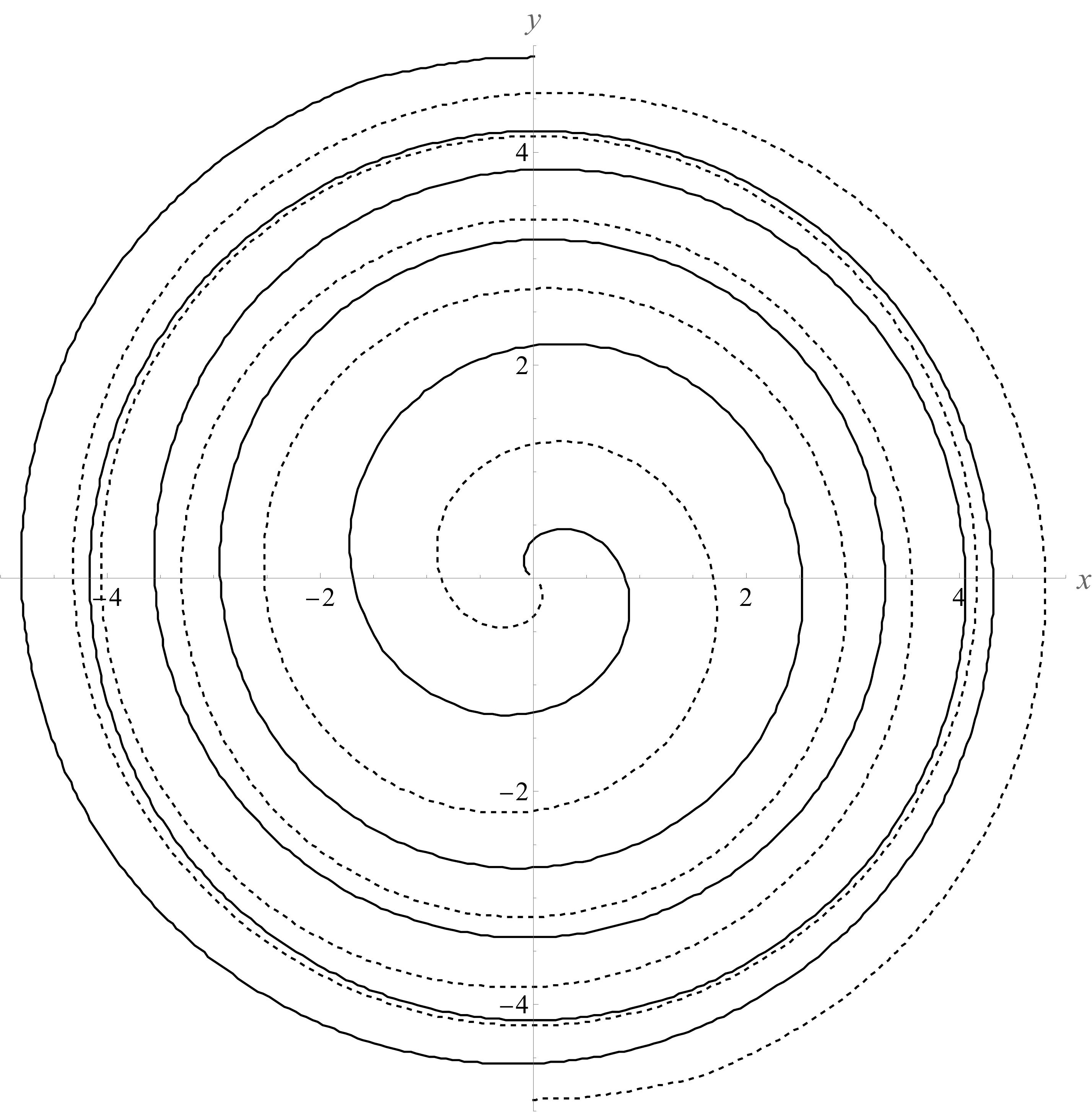}
	\caption{Blackhole binary trajectories.  The black hole "centers" are tracked by following the minimized lapse (solid and dashed lines denote each black hole center in the trajectory).}
	\label{BHTrajectory}
\end{figure}

Figure \ref{BHTrajectory} illustrates the trajectory of the black holes as the simulation progresses.  The results are similar to trajectories for equal mass binaries as obtained by previous analysis.  Though on should be aware that differences in gauge conditioning and evolution schemes can lead to quantitative differences in the plotted trajectories.  

\begin{figure}
	\centering
	\includegraphics[width=0.9\linewidth]{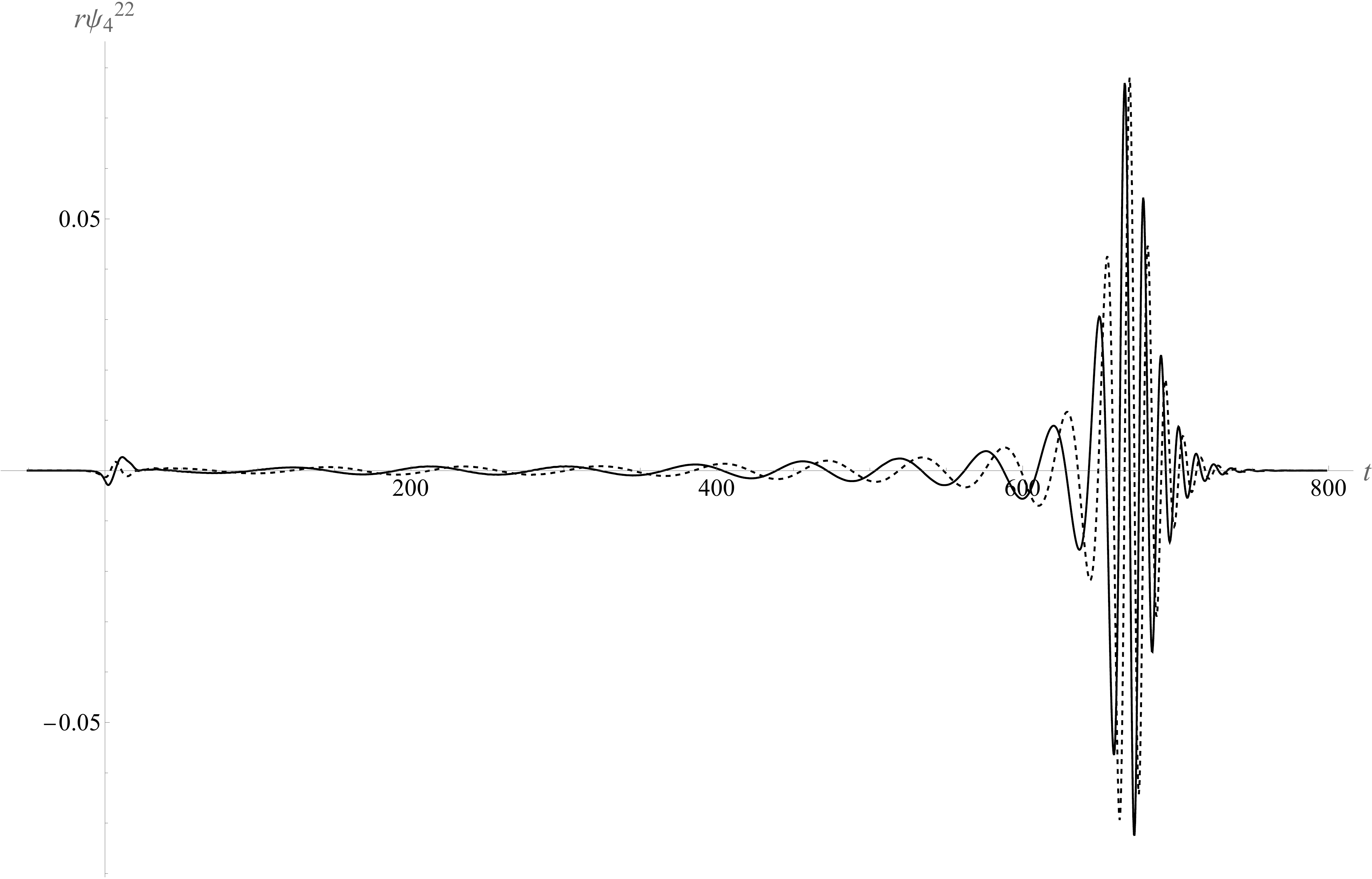}
	\caption{Blackhole binary $(l=2,\; m = 2)$ waveforms for both the real (solid) and imaginary (dashed) parts of $r\psi_4^{22}$.}
	\label{BHSpiralWaveForms}
\end{figure}

As in the previous case of the head on collision, the simulation is analyzed by extracting the wave forms at fixed radii in the linear regime.  Here we extract the wave forms at coordinate radius $r = 50$ outside the FMR region (defined for $r > 32$, where regridding does not take place).  In this case we calculate the $l=2, m=2$ spherical harmonic amplitude of the Weyl scalar $\psi_4^{22}$.  Figure \ref{BHSpiralWaveForms} illustrates the time dependent values of the spherical wave forms at the extraction radius.

\section{Summary and Discussion}

To summarize our presentation in this paper, we have developed and demonstrated a code generation method for constructing complex PDE solvers for AMReX applications.  The method of code generation significantly reduces the time spent analyzing and debugging large lines of code.  Additionally code generation allows one to generate executable code for low level (fast) languages from symbolic tensor manipulation packages in higher level languages found in python and Mathematica applications. 

We have demonstrated the application of code generation to the complex equations involved in a $3+1$ formulation of numerical relativity.  In particular we have adapted our code generator to produce executables for AMReX applications of numerical relativity.  We find that the generated systems of equations for Z4c and CCZ4 formulations produce accurate results when probed with the AwA tests for numerical relativity solvers.  We thus conclude that the generated code is reliable and accurate, and may be used in more general contexts beyond the construction of numerical relativity solvers.

As an example of future applications, we intend to adapt the code generator to consider more physically realistic spacetime systems such as those of neutron star binaries, or core collapse supernovae.  Such endevours will require implementation of proper boundary conditions, such as Sommerfeld conditions for outward propagating waves, which will further require the use of code generation within AMReX. Such applications will further push the developement of code generation as one considered hydrodynamics and magnetohydrodynamics in a dynamical spacetime.  Additionally, code generation opens the door for consideration of more exotic spacetime solves such as those associated with time evolution in holographic spacetime problems such as those encountered in the AdS/CFT correspondence, or holographic superconductors. 

\section*{Acknowledgements}
This research was funded by the ExaStar project within the Exascale Computing Project (17-SC-20-SC).  Large scale computations were performed using the Cori supercomputer of the National Energy Research Scientific Computing Center.

AP is very thankful to Erik Schnetter, Vasilli Mewes, and Ann Almgren for useful discussions and advice for this project.

\section{Appendices}
\subsection*{Appendix A: Review of Bowen York initial data}
Here we review the method of generating black hole binary initial data as presented in \cite{Bowen:1980yu} and \cite{Kulkarni:1983rsa}.  The initial data are generated by ensuring that the constraints (\ref{PhysicalConstraints}) are satisfied on an initial spatial hypersurface $\Sigma$ of the spacetime manifold.  Following the original conformal decomposition of the spatial metric 
\begin{equation}
	\gamma_{ij} = \psi^4\tilde{\gamma}_{ij},
	\label{ConformalDecompositionGamma}
\end{equation}
where we switch from $\phi$ and $\chi$ to $\psi$ as the conformal factor for convenience.  We may then insert (\ref{ConformalDecompositionGamma}) into the Hamiltonian constraint (\ref{PhysicalConstraints}):
\begin{equation}
	8\tilde{D}^2 \psi - \psi R - \psi^5 K^2 + \psi^{-7}\hat{A}^{ij}\hat{A}_{ij} = 0,
	\label{HamiltonianConstraint}
\end{equation}
where we have introduced the conformal decomposition of $\hat{A}_{ij} \equiv \psi^2 A_{ij}$.  Hear $\tilde{D}^2 = \tilde{\gamma}^{ij} \tilde{D}_i \tilde{D}_j$ is the covariant Laplacian with $\tilde{D}_i$ the covariant derivative compatible with the spatial conformal metric $\tilde{\gamma}_{ij}$.

With these definitions we rewrite the momentum constraint (\ref{PhysicalConstraints}) as:
\begin{equation}
	\tilde{D}_j \hat{A}^{ij} - \frac{2}{3}\psi^6 \tilde{\gamma}_{ij}\tilde{D}_j K = 0.
	\label{MomentumConstraintInitialData}
\end{equation}

Analytical solutions to equation (\ref{MomentumConstraintInitialData}) compatible with black hole spacetimes are well known. In particular we may chose the conformal transverse traceless approach with an initally flat conformal metric $\tilde{\gamma}_{ij} = \eta_{ij}$ with maximal slicing $K=0$ (Bowen-York initial data).  In this case one is left with
\begin{equation}
	\partial_j \hat{A}^{ij} = 0
	\label{MomentumConstraintSimple}
\end{equation}
which is linear, and thus possess simple solutions.  Specifically, one may consider a single moving black hole with initial position $C^i$, and initial momentum $P^i$ in Cartesian coordinates.  Solutions possessing these characteristics may be written as:
\begin{equation}
	\hat{A}_{\bf CP}^{ij} = \frac{2}{3 r_{\bf C}^2}\left(P^i n_{\bf C}^j + P^j n_{\bf C}^i -(\eta^{ij}-n_{\bf C}^i n_{\bf C}^j)n_{\bf C}^k P_k\right).
	\label{MomentumConstraintSolution}
\end{equation}
The linearity of (\ref{MomentumConstraintSimple}) allows one to add several solutions of (\ref{MomentumConstraintSolution}) together.  In particular one may select several values for ${\bf C}$ and ${\bf P}$ and write a general solution
\begin{equation}
	\hat{A}^{ij} = \sum_{n} \hat{A}_{{\bf C}_n{\bf P}_n }^{ij}.
\end{equation}

We then turn to the Hamiltonian constraint (\ref{HamiltonianConstraint}) with conformal flatness and maximal slicing:
\begin{equation}
	\nabla^2 \psi = -\frac{1}{8} \psi^{-7} \hat{A}_{ij}\hat{A}^{ij}.
	\label{HamiltonianConstraintSimple}
\end{equation}
In the puncture approach to solving (\ref{HamiltonianConstraintSimple}) one absorbs the analytic singularities into $\psi$ and considers corrections to $\psi$ that are solved numerically.  To this end for a black hole binary we write:
\begin{equation}
	\psi = 1 + \frac{1}{\xi} + u,
\label{psidecomposition}
\end{equation}
for freely chosen bare masses $M_{1,2}$.  Here $\xi$ is defined as:
\begin{equation}
	\xi \equiv \frac{M_1}{2r_{{\bf C}_1}} + \frac{M_2}{2r_{{\bf C}_2}}.
	\label{xidefinition}
\end{equation}  
In this case the Hamiltonian constraint (\ref{HamiltonianConstraintSimple}) reduces to:
\begin{equation}
	\nabla^2 u = -\frac{1}{8}\hat{A}_{ij}\hat{A}^{ij} \left(\frac{\xi}{\xi (1+u)+1}\right)^7.
	\label{HamiltonianConstraintSimpleU}
\end{equation}
One may then employ standard methods of numerical solutions to elliptic equations to find $u$.  For this particular project we employed a simple relaxation procedure where the initial guess for $u$ was determined via the approximation described in \cite{Dennison:2006nq}.

The grid setup and FMR/AMR algorithm for the initial data is determined by the grid setup of the problem we intend to evolve.  Grid and AMR inputs are chosen to match for both initial data and Z4c (or CCZ4) solver.  This allows for easy copying from one solver to the next.

\subsection*{Appendix B: Conformal Covariant Z4 (CCZ4) system}

For certain tests and physical situations it is advantageous to consider an alternate to the Z4c formulation known as the conformal covariant Z4 (CCZ4) formulation for evolution of the equations of motion.  The CCZ4 is based on the same conformal decomposition of the original Z4 system as shown in (\ref{Z4cEE}), (\ref{Z4gammaK}) and (\ref{Z4ThetaZ}).  The CCZ4 amounts to a rearrangement of terms in the Z4 system with additional adjustable parameters determining the conformal covariance of the evolution of the equations of motion.  For particular tests such as the gauge wave test, it has been well documented that the constraint damped CCZ4 leads to improved stability of the evolution.  However, the desirability of conformal covariance presents stability issues with black hole binary evolutions.  In these cases the conformal covariance of the equations is sacrificed for numerical stability.  

We present the final equations of the CCZ4 system with the variable definitions given in section 2.  We however adjust our definition of the conformal factor in the spatial metric as:
\begin{equation}
	\tilde{\gamma}_{ij} = W^2\gamma_{ij},
\end{equation}
and thus we have a redefinition of the conformal variable $W = \exp(-2\phi)$.  See \cite{Marronetti:2007wz} and \cite{Cao:2008wn} for details and justification for this redefinition.

With this redefinition the CCZ4 system decomposes as:

\begin{align}
	\partial_t W &= \frac{1}{3}\alpha W K + \beta^{i}\partial_{i} W -\frac{1}{3} W \partial_i \beta^i \\
	\partial_t \tilde{\gamma}_{ij} &= -2 \alpha \tilde{A}_{ij} + 2 \tilde{\gamma}_{k(i}\partial_{j)}\beta^k -\frac{2}{3}\tilde{\gamma}_{ij}\partial_k \beta^k+ \beta^k \partial_k\tilde{\gamma}_{ij} \\
	\partial_t K &= - D_i D^i \alpha + \alpha \left( R + 2D_iZ^i + K^2 -2 \Theta K\right) - 3\alpha \kappa_1 (1+\kappa_2)\Theta \nonumber \\
	&+ 4\pi\alpha(S-3\rho_{\rm ADM}) + \beta^i\partial_i K\\
	\partial_t \tilde{A}_{ij} &= W^2\left[ -D_i D_j \alpha + \alpha \left( R_{ij} + 2 D_{(i} Z_{j)} - 8 \pi S_{ij}\right)\right]^{\rm tf} + \alpha \left[ \left( K - 2\Theta \right) \tilde{A}_{ij} \right. \nonumber \\
	&\left. \;\;\; -2\tilde{A}_{ik}\tilde{A}^k_j\right] + 2\tilde{A}_{k (i}\partial_{j)}\beta^k-\frac{2}{3}\tilde{A}_{ij} \partial_k \beta^k+\beta^k \partial_k \tilde{A}_{ij} \\
	\partial_t \Theta &= \frac{1}{2}\alpha\left[R + 2 D_i Z^i - \tilde{A}_{ij}\tilde{A}^{ij} + \frac{2}{3}K^2 - 2 \Theta K - 16 \pi \rho_{\rm ADM} \right. \nonumber \\
	& \left. \phantom{\frac{0}{0}}\;\; -2\kappa_1 (2+\kappa_2)\Theta \right] -Z^i\partial_i \alpha + \beta^i \partial_i \Theta \\
	\partial_t \tilde{\Gamma}^i &= \tilde{\gamma}^{jk} \partial_j \partial_k \beta^i + \frac{1}{3}\tilde{\gamma}^{ij}\partial_j \partial_k \beta^k - 2\tilde{A}^{ij} \partial_j \alpha + 2\alpha \left[ \tilde{\Gamma}^i_{jk} \tilde{A}^{jk} - 3\tilde{A}^{ij}\frac{\partial_j W}{W} \right. \nonumber \\
	& \left. \;\;\;-\frac{2}{3} \tilde{\gamma}^{ij}\partial_j K - 8\pi\tilde{\gamma}^{ij}S_j \right] + 2\tilde{\gamma}^{ik}\left[ \alpha \partial_k \Theta - \Theta \partial_k \alpha -\frac{2}{3}\alpha K Z_k \right] \nonumber \\
	&\;\;\;+ \frac{2}{3} \tilde{\Gamma}_{\rm d}^i \partial_j \beta^j - \tilde{\Gamma}_{\rm d}^j\partial_j \beta^i + 2\kappa_3\left[\frac{2}{3}\tilde{\gamma}^{ij}Z_j \partial_k \beta^k -\tilde{\gamma}^{jk}Z_j\partial_k\beta^i\right] \nonumber \\
	& \;\;\; -2\alpha \kappa_1\tilde{\gamma}^{ij}Z_j+ \beta^j\partial_j \tilde{\Gamma}^i.
	\label{CCZ4}
\end{align}	
Here the Ricci tensor $R_{ij}$ of $\gamma_{ij}$ can be decomposed into a conformal and the Ricci tensor $\tilde{R}_{ij}$ associated with $\tilde{\gamma}_{ij}$.  Additionally, the constraints associated with $Z_i$ are absorbed into the definition of $\tilde{R}_{ij}$:
\begin{align}
	R_{ij} &= R^{W}_{ij} + \tilde{R}_{ij} \\
	R^{W}_{ij} &= \frac{1}{W^2}\left[ W \left(\tilde{D}_i \tilde{D}_j W + \tilde{\gamma}_{ij} \tilde{D}_l \tilde{D}^l W\right) - 2\tilde{\gamma}_{ij } \tilde{D}^l W \tilde{D}_l W\right] \\
	\tilde{R}_{ij} &= -\frac{1}{2}\tilde{\gamma}^{lm}\partial_i \partial_j \tilde{\gamma}_{lm} + \tilde{\gamma}_{k(i}\partial_{j)}\tilde{\Gamma}_{\rm d}^k + \tilde{\Gamma}_{\rm d}^k\tilde{\Gamma}_{(ij)k} \nonumber \\
	& \;\;\;\; +\tilde{\gamma}^{lm}\left(2\tilde{\Gamma}^k_{l(i}\tilde{\Gamma}_{j)km} + \tilde{\Gamma}^k_{im} \tilde{\Gamma}_{klj}\right).
\label{CCZ4R}
\end{align}
Note the slight difference in definition of $\tilde{R}_{ij}$ in the second term with that in the Z4c formulation in (\ref{Z4cR}).  As in the Z4c formulation we enforce the algebraic constraints as in (\ref{AlgbraicConstraint}).

Here the puncture gauge conditions for the lapse $\alpha$, and shift $\beta^i$ take the form:
\begin{align}
	\partial_t \alpha &= -\mu_{\rm L} \alpha^2 (K - 2\Theta) + \beta^i \partial_i \alpha \\
	\partial_t \beta^i &= \mu_{\rm S} \alpha^2 \tilde{\Gamma}^i - \eta \beta^i + \beta^j \partial_j \beta^i.
	\label{gaugeCCZ4}
\end{align}

\subsection*{Appendix C: Newman-Penrose waveform extraction}

To extract the gravitational waveform of a black hole collision, we adopt the Newman-Penrose formulation (see \cite{Campanelli:1997} \cite{Campanelli:1998uh} \cite{Campanelli:1998yt} \cite{Campanelli:1998jv} \cite{Baker:2001sf} \cite{Campanelli:2005ia}).  Here the 10 components of the Weyl tensor $^{(4)}C_{\mu\nu\alpha\beta}$ is decomposed into 5 complex valued scalars constructed from contractions with a null tetrad $(l^\alpha,k^\alpha, m^\alpha, \bar{m}^\alpha)$.  Considering quasi-Kinnersly tetrads it is possible to show that the Weyl scalar $\psi_4$ defined as
\begin{equation}
	\psi_4 \equiv -^{(4)}C_{\mu\nu\alpha\beta} k^\mu \bar{m}^\nu k^\alpha m^\beta,
\end{equation} 
has the property that as $r \rightarrow \infty$,
\begin{equation}
	\psi_4 \rightarrow \partial_t^2 h_+ - i \partial_t^2 h_\times,
\end{equation}
where $h_+$ and $h_\times$ are the two independent transverse modes of the metric perturbation.

Performing the $3+1$ decomposition it can be shown:
\begin{align}
	\psi_4 &= (R_{ijkl} + 2K_{i[k}K_{l]j})n^i\bar{m}^j n^k \bar{m}^l \nonumber \\
	& \;\;\;-8(K_{j[k,l]} + \Gamma^p_{j[k}K_{l]p})n^{[0}\bar{m}^{j]}n^k \bar{m}^l \nonumber \\
	& \;\;\;+4(R_{jl}-K_{jp}K^p_l+K K_{jl})n^{[0}\bar{m}^{j]}n^{[0}\bar{m}^{l]}.
\label{WeylScalarFinalForm}
\end{align}

Following the general expression for the Weyl scalar in (\ref{WeylScalarFinalForm}), we implement the quasi-Kinnersley tetrad as constructed in [Baker, Campanelli, Lousto].  In this procedure an orthogonal set of spatial vectors aligned with the $\hat(\varphi)$ and $\hat(r)$ directions is selected, along with a third vector constructed from the cross product.  These are written in Cartesian coordinates as:
\begin{align}
    &v_1^a = [-y, x, 0], \nonumber \\
    &v_2^a = [x, y, z], \nonumber \\
    &v_3^a = \det(\gamma)^{1/2} \gamma^{ad}\epsilon_{dbc}v_1^b v_2^c,
\end{align}
where $\epsilon_{abc}$ is the standard Levi-Cevita symbol with $\epsilon_{123} = 1$.

An orthonormal basis is then constructed via the Gram-Schmidt procedure.  Specifically, orthonormal vectors $v_1^a, v_2^a,$ and $v_3^a$ are manipulated one after the other as follows:
\begin{align}
    v_1^a &\rightarrow \frac{v_1^a}{\sqrt{\omega_{11}}}, \nonumber \\
    v_2^a &\rightarrow \frac{v_2^a-\omega_{12}}{\sqrt{\omega_{22}}}, \nonumber \\
    v_3^a &\rightarrow \frac{v_3^a-\omega_{13}v_1^a - \omega_{23}v_2^a}{\sqrt{\omega_{33}}},
\label{GSOrthoBasis}
\end{align}
where $\omega_{ij} \equiv v_i^a v_j^b \gamma_{ab}$.  Note that for each $v_i^a$ in (\ref{GSOrthoBasis}), the numerator is evaluated first, and then the denominator with newly constructed value of $v_i^a$.

The tetrad is now constructed using the orthonormal basis (\ref{GSOrthoBasis}) along with the time-like unit normal $u^\mu$, which in the 3+1 decomposition takes the form:
\begin{equation}
    u^\mu = \frac{1}{\alpha}\left(1, -\beta^i\right).
\end{equation}
The null tetrad is then constructed as:
\begin{align}
    l^\mu &= \frac{1}{\sqrt{2}}\left(u^\mu + r^\mu\right) \nonumber \\
    n^\mu &= \frac{1}{\sqrt{2}}\left(u^\mu - r^\mu\right) \nonumber \\
    m^\mu &= \frac{1}{\sqrt{2}}\left(\theta^\mu + i\varphi^\mu\right).
\end{align}

The Weyl scalar transforms as a spherical tensor with spin weight $-2$ and is thus decomposed with the spin weighted spherical harmonics $Y^{-2}_{lm}$:
\begin{equation}
    \psi_4(t,r,\theta, \varphi) = \sum_{l=2}^\infty \sum_{m=-l}^{m = l} \psi_4^{lm}(t,r)Y^{-2}_{lm}(\theta,\varphi).
\end{equation}
with spherical amplitude:
\begin{equation}
    \psi_4^{lm} = \int d\Omega Y^{-2\star}_{lm} \psi_4.
\end{equation}

The spin-weighted spherical harmonics are defined as:
\begin{equation}
    Y^{-2}_{lm} \equiv \sqrt{\frac{(l-2)!}{(l+2)!}}\left(W_{lm}(\theta, \varphi) - i\frac{X_{lm}(\theta,\varphi)}{\sin\theta}\right),
\end{equation}
with $W_{lm}(\theta,\varphi)$ and $X_{lm}(\theta,\varphi)$ defined as:
\begin{align}
    W_{lm}(\theta,\varphi) &= \left(\partial_\theta^2 - \cot\theta \partial_\theta - \frac{\partial_\varphi^2}{\sin^2\theta}\right) Y_{lm}(\theta,\varphi), \nonumber \\
    X_{lm} &= 2\partial_\varphi \left(\partial_\theta - \cot\theta\right) Y_{lm}(\theta,\varphi),
\end{align}
and the standard definition of the spherical harmonics:
\begin{equation}
    Y_{lm}(\theta,\varphi) = \sqrt{\frac{2l+1}{4 \pi}}\sqrt{\frac{(l-m)!}{(l+m)!}} P^l_m(\cos\theta)e^{i m \varphi},
\end{equation}
where $P_m^l$ are the associated Legendre polynomials.

For the cases we consider for head on and inspiraling black hole binary collisions we explicitly use the $s=-2$ weighted spherical harmonics:
\begin{align}
    Y^{-2}_{20} &= \frac{3}{4}\sqrt{\frac{5}{6\pi}}\sin^2\theta, \nonumber \\
    Y^{-2}_{22} &= \frac{1}{8}\sqrt{\frac{5}{\pi}}\left(1+\cos\theta\right)^2 e^{2 i \pi}.
\end{align}


\begin{thebibliography}{99}

\bibitem{Abramovici:1992ah}
A.~Abramovici, W.~E.~Althouse, R.~W.~P.~Drever, Y.~Gursel, S.~Kawamura, F.~J.~Raab, D.~Shoemaker, L.~Sievers, R.~E.~Spero and K.~S.~Thorne, \textit{et al.}
Science \textbf{256}, 325-333 (1992)
doi:10.1126/science.256.5055.325

\bibitem{Caron:1997hu}
B.~Caron, A.~Dominjon, C.~Drezen, R.~Flaminio, X.~Grave, F.~Marion, L.~Massonnet, C.~Mehmel, R.~Morand and B.~Mours, \textit{et al.}
Class. Quant. Grav. \textbf{14}, 1461-1469 (1997)
doi:10.1088/0264-9381/14/6/011

\bibitem{LIGOScientific:2016aoc}
B.~P.~Abbott \textit{et al.} [LIGO Scientific and Virgo],
Phys. Rev. Lett. \textbf{116}, no.6, 061102 (2016)
doi:10.1103/PhysRevLett.116.061102
[arXiv:1602.03837 [gr-qc]].

\bibitem{Baker:2005vv}
J.~G.~Baker, J.~Centrella, D.~I.~Choi, M.~Koppitz and J.~van Meter,
Phys. Rev. Lett. \textbf{96}, 111102 (2006)
doi:10.1103/PhysRevLett.96.111102
[arXiv:gr-qc/0511103 [gr-qc]].

\bibitem{Campanelli:2005dd}
M.~Campanelli, C.~O.~Lousto, P.~Marronetti and Y.~Zlochower,
Phys. Rev. Lett. \textbf{96}, 111101 (2006)
doi:10.1103/PhysRevLett.96.111101
[arXiv:gr-qc/0511048 [gr-qc]].

\bibitem{Gundlach:2006tw}
C.~Gundlach and J.~M.~Martin-Garcia,
Phys. Rev. D \textbf{74}, 024016 (2006)
doi:10.1103/PhysRevD.74.024016
[arXiv:gr-qc/0604035 [gr-qc]].

\bibitem{Bona:2003fj}
C.~Bona, T.~Ledvinka, C.~Palenzuela and M.~Zacek,
Phys. Rev. D \textbf{67}, 104005 (2003)
doi:10.1103/PhysRevD.67.104005
[arXiv:gr-qc/0302083 [gr-qc]].

\bibitem{Gundlach:2005eh}
C.~Gundlach, J.~M.~Martin-Garcia, G.~Calabrese and I.~Hinder,
Class. Quant. Grav. \textbf{22}, 3767-3774 (2005)
doi:10.1088/0264-9381/22/17/025
[arXiv:gr-qc/0504114 [gr-qc]].

\bibitem{Bernuzzi:2009ex}
S.~Bernuzzi and D.~Hilditch,
Phys. Rev. D \textbf{81}, 084003 (2010)
doi:10.1103/PhysRevD.81.084003
[arXiv:0912.2920 [gr-qc]].

\bibitem{Babiuc:2007vr}
M.~C.~Babiuc, S.~Husa, D.~Alic, I.~Hinder, C.~Lechner, E.~Schnetter, B.~Szilagyi, Y.~Zlochower, N.~Dorband and D.~Pollney, \textit{et al.}
Class. Quant. Grav. \textbf{25}, 125012 (2008)
doi:10.1088/0264-9381/25/12/125012
[arXiv:0709.3559 [gr-qc]].

\bibitem{Husa:2004ip}
S.~Husa, I.~Hinder and C.~Lechner,
Comput. Phys. Commun. \textbf{174}, 983-1004 (2006)
doi:10.1016/j.cpc.2006.02.002
[arXiv:gr-qc/0404023 [gr-qc]].

\bibitem{Ruchlin:2017com}
I.~Ruchlin, Z.~B.~Etienne and T.~W.~Baumgarte,
Phys. Rev. D \textbf{97}, no.6, 064036 (2018)
doi:10.1103/PhysRevD.97.064036
[arXiv:1712.07658 [gr-qc]].

\bibitem{Bona:1994dr}
C.~Bona, J.~Masso, E.~Seidel and J.~Stela,
Phys. Rev. Lett. \textbf{75}, 600-603 (1995)
doi:10.1103/PhysRevLett.75.600
[arXiv:gr-qc/9412071 [gr-qc]].

\bibitem{McCorquodale:2011}
P.~McCorquodale and P. Colella,
Comm. App. Math. and Comp. Sci. \textbf{6}, no.1, 1-25 (2011)
doi.10.2140/camcos.2011.6.1

\bibitem{Alcubierre:2002kk}
M.~Alcubierre, B.~Bruegmann, P.~Diener, M.~Koppitz, D.~Pollney, E.~Seidel and R.~Takahashi,
Phys. Rev. D \textbf{67}, 084023 (2003)
doi:10.1103/PhysRevD.67.084023
[arXiv:gr-qc/0206072 [gr-qc]].

\bibitem{Brill:1963yv}
D.~R.~Brill and R.~W.~Lindquist,
Phys. Rev. \textbf{131}, 471-476 (1963)
doi:10.1103/PhysRev.131.471

\bibitem{Brandt:1997xd}
S.~R.~Brandt and B.~Bruegmann,
[arXiv:gr-qc/9711015 [gr-qc]].

\bibitem{Okawa:2013afa}
H.~Okawa,
Int. J. Mod. Phys. A \textbf{28}, 1340016 (2013)
doi:10.1142/S0217751X13400162
[arXiv:1308.3502 [gr-qc]].

\bibitem{Bowen:1980yu}
J.~M.~Bowen and J.~W.~York, Jr.,
Phys. Rev. D \textbf{21}, 2047-2056 (1980)
doi:10.1103/PhysRevD.21.2047

\bibitem{Kulkarni:1983rsa}
A.~Kulkarni, L.~Shepley and J.~W.~York, Jr.,
Phys. Lett. A \textbf{96}, 228-230 (1983)
doi:10.1016/0375-9601(83)90338-9

\bibitem{Cao:2011fu}
Z.~Cao and D.~Hilditch,
Phys. Rev. D \textbf{85}, 124032 (2012)
doi:10.1103/PhysRevD.85.124032
[arXiv:1111.2177 [gr-qc]].

\bibitem{Dennison:2006nq}
K.~A.~Dennison, T.~W.~Baumgarte and H.~P.~Pfeiffer,
Phys. Rev. D \textbf{74}, 064016 (2006)
doi:10.1103/PhysRevD.74.064016
[arXiv:gr-qc/0606037 [gr-qc]].

\bibitem{Marronetti:2007wz}
P.~Marronetti, W.~Tichy, B.~Bruegmann, J.~Gonzalez and U.~Sperhake,
Phys. Rev. D \textbf{77}, 064010 (2008)
doi:10.1103/PhysRevD.77.064010
[arXiv:0709.2160 [gr-qc]].

\bibitem{Cao:2008wn}
Z.~j.~Cao, H.~J.~Yo and J.~P.~Yu,
Phys. Rev. D \textbf{78}, 124011 (2008)
doi:10.1103/PhysRevD.78.124011
[arXiv:0812.0641 [gr-qc]].

\bibitem{Campanelli:1997}
M.~Campanelli, W.~Krivan and C.~O.~Lousto,
Phys. Rev. D \textbf{58}, 024016 (1998)
doi:10.1103/PhysRevD.58.024015

\bibitem{Campanelli:1998uh}
M.~Campanelli, W.~Krivan and C.~O.~Lousto,
Phys. Rev. D \textbf{58}, 024016 (1998)
doi:10.1103/PhysRevD.58.024016
[arXiv:gr-qc/9801067 [gr-qc]].

\bibitem{Campanelli:1998yt}
M.~Campanelli, C.~O.~Lousto, J.~G.~Baker, G.~Khanna and J.~Pullin,
Phys. Rev. D \textbf{58}, 084019 (1998)
[erratum: Phys. Rev. D \textbf{62}, 069901 (2000)]
doi:10.1103/PhysRevD.58.084019
[arXiv:gr-qc/9803058 [gr-qc]].

\bibitem{Campanelli:1998jv}
M.~Campanelli and C.~O.~Lousto,
Phys. Rev. D \textbf{59}, 124022 (1999)
doi:10.1103/PhysRevD.59.124022
[arXiv:gr-qc/9811019 [gr-qc]].

\bibitem{Baker:2001sf}
J.~G.~Baker, M.~Campanelli and C.~O.~Lousto,
Phys. Rev. D \textbf{65}, 044001 (2002)
doi:10.1103/PhysRevD.65.044001
[arXiv:gr-qc/0104063 [gr-qc]].

\bibitem{Campanelli:2005ia}
M.~Campanelli, B.~J.~Kelly and C.~O.~Lousto,
Phys. Rev. D \textbf{73}, 064005 (2006)
doi:10.1103/PhysRevD.73.064005
[arXiv:gr-qc/0510122 [gr-qc]].



\end{thebibliography}
\end{document}